\documentclass[
 preprint,
superscriptaddress,
nofootinbib,
 amssymb,
 aps,
 prd,
 onecolumn
]{revtex4-2}

\usepackage{graphicx}%
\usepackage{dcolumn}%
\usepackage{bm}%
\usepackage{verbatim}%
\usepackage{hyperref}%

\usepackage{ragged2e} %
\usepackage[caption=false, margin=0.02\textwidth]{subfig}
\usepackage[export]{adjustbox}

\makeatletter
\renewcommand{\frontmatter@RRAPformat}[1]{
\begin{center}

\vspace{10pt}
   \expandafter\produce@RRAP\expandafter{\@date}%
   \expandafter\produce@RRAP\expandafter{\@received}%
   \expandafter\produce@RRAP\expandafter{\@revised}%
   \expandafter\produce@RRAP\expandafter{\@accepted}%
   \expandafter\produce@RRAP\expandafter{\@published}%
\end{center}
}
\makeatother

\makeatletter
\renewcommand\@makecaption[2]{%
  \vskip\abovecaptionskip
  \justifying
  \sbox\@tempboxa{#1: #2}%
  \ifdim \wd\@tempboxa > \hsize
    \noindent #1: #2\par
  \else
    \hbox to\hsize{\hfil\box\@tempboxa\hfil}%
  \fi
  \vskip\belowcaptionskip}
\makeatother

\makeatletter
\renewcommand{\fnum@figure}{Figure~\thefigure}
\makeatother

\makeatletter
\renewcommand{\fnum@table}{Table~\thetable}
\makeatother
\renewcommand{\thetable}{\arabic{table}}

\usepackage{placeins}
\usepackage{siunitx}%
\usepackage{xspace}

\usepackage[dvipsnames]{xcolor}

\usepackage[normalem]{ulem}
\def\mumu{\ensuremath{\mu^+\mu^-}~}
\def\ee{\ensuremath{e^+e^-}~}

\newcommand{\pT}{\ensuremath{p_{\text{T}}}\xspace}

\begin{document}

\title{MAIA: A new detector concept for a 10 TeV muon collider}%

\author{Charles Bell}\affiliation{University of Tennessee, Knoxville, TN, USA}
\author{Daniele Calzolari}\affiliation{European Organization for Nuclear Research (CERN), Switzerland}
\author{Christian Carli}\affiliation{European Organization for Nuclear Research (CERN), Switzerland}
\author{John Dervan}\affiliation{University of Tennessee, Knoxville, TN, USA}
\author{Karri Folan Di Petrillo}\affiliation{University of Chicago, IL, USA}
\author{Micah Hillman}\affiliation{University of Tennessee, Knoxville, TN, USA}
\author{Tova R. Holmes}\affiliation{University of Tennessee, Knoxville, TN, USA}
\author{Sergo Jindariani}\affiliation{Fermi National Accelerator Laboratory, IL, USA}
\author{Kiley E. Kennedy}\affiliation{Princeton University, NJ, USA}
\author{Cyrus Kianian}\affiliation{University of Wisconsin–Madison, WI, USA}
\author{Ka Hei Martin Kwok}\affiliation{Fermi National Accelerator Laboratory, IL, USA} \affiliation{University of Nebraska–Lincoln, Lincoln, NE, USA}
\author{Mark Larson}\affiliation{University of Chicago, IL, USA}
\author{Anton Lechner}\affiliation{European Organization for Nuclear Research (CERN), Switzerland}
\author{Lawrence Lee}\affiliation{University of Tennessee, Knoxville, TN, USA}
\author{Thomas Madlener} \affiliation{Deutsches Elektronen-Synchrotron DESY, Germany}
\author{Federico Meloni} \affiliation{Deutsches Elektronen-Synchrotron DESY, Germany}
\author{Abdollah Mohammadi}\affiliation{University of Wisconsin–Madison, WI, USA}
\author{Isobel Ojalvo}\affiliation{Princeton University, NJ, USA}
\author{Priscilla Pani} \affiliation{Deutsches Elektronen-Synchrotron DESY, Germany}
\author{Gregory Douglas Penn}\affiliation{Yale University, CT, USA}
\author{Rose Powers}\affiliation{Princeton University, NJ, USA}
\author{Benjamin Rosser}\affiliation{University of Chicago, IL, USA}
\author{Leo Rozanov}\affiliation{University of Chicago, IL, USA} \affiliation{Imperial College London, UK}
\author{Kyriacos Skoufaris}\affiliation{European Organization for Nuclear Research (CERN), Switzerland}
\author{Elise Sledge}\affiliation{Princeton University, NJ, USA} \affiliation{California Institute of Technology, CA, USA}
\author{Alexander Tuna}\affiliation{University of Tennessee, Knoxville, TN, USA}
\author{Junjia Zhang}\affiliation{Princeton University, NJ, USA}

\date{\today}%

\begin{abstract}
Muon colliders offer a compelling opportunity to explore the TeV scale and conduct precision tests of the Standard Model, all within a relatively compact geographical footprint. This paper introduces a new detector concept, MAIA (Muon Accelerator Instrumented Apparatus), optimized for $\sqrt{s}=10$~TeV \mumu collisions. The detector features an all-silicon tracker immersed in a 5T solenoid field. High-granularity silicon-tungsten and iron-scintillator calorimeters surrounding the solenoid capture high-energy electronic and hadronic showers, respectively, and support particle-flow reconstruction. 
The outermost subsystem comprises an air-gap muon spectrometer, which contributes to muon identification.
The performance of the MAIA detector is evaluated in terms of differential particle reconstruction efficiencies and resolutions. Beam-induced background and incoherent pair production simulations are overlaid to single particle gun samples to assess detector reconstruction capabilities under realistic experimental conditions. Even in the presence of backgrounds, reconstruction efficiencies exceed approximately 95\% for energetic tracks, photons, and charged pions in the central region of the detector. This paper outlines promising avenues for future work, including forward region optimization, opportunities for enhanced flavor tagging and boosted object reconstruction, and technological developments needed to achieve the desired detector performance.
\end{abstract}

\maketitle

\clearpage
\tableofcontents

\newpage

\section{Introduction}
Particle colliders have played a central role in experimentally testing the Standard Model (SM) of particle physics, from the discovery of the charm quark in 1974 to that of the Higgs boson in 2012~\cite{CharmDiscoverySLAC,CharmDiscoveryE598,HiggsDiscoveryATLAS,HiggsDiscoveryCMS}. Over the past half-century, the hunt for new physics in collider experiments has alternated between two complementary approaches: either direct searches for particles with higher masses or smaller couplings, or precision measurements of SM parameters to probe new physics indirectly. While the Large Hadron Collider (LHC) will remain at the forefront of the energy frontier of particle physics for the next twenty years, a higher energy collider will be instrumental to further our understanding of fundamental particle physics in the following decades. 

Muon colliders are an appealing option for future particle accelerators due to their ability to reach high energies with a relatively compact size, high power efficiency, and low cost~\cite{Roser_2023}. Extensive studies on the physics potential of multi-TeV muon colliders suggest that these machines offer competitive sensitivity to both the precision measurements of \ee machines and the discovery potential of pp colliders with comparable parton center-of-mass energies~\cite{InternationalMuonCollider:2025sys,accettura2023muoncollider}. Various staging options for a TeV-scale muon collider have been proposed~\cite{accettura2023muoncollider,delahaye2019muoncolliders,Long:2020wfp,
delahaye2015stagedmuonacceleratorfacility}. In one scenario currently being considered by the International Muon Collider Collaboration (IMCC), an initial 3~TeV machine would be followed by a 10~TeV collider; in other scenarios, a 10~TeV machine would be built immediately but with run with lower luminosity to be upgraded over time~\cite{InternationalMuonCollider:2025sys}. In both cases, reaching the 10~TeV energy scale offers an unprecedented opportunity for both discovery and precision.

The primary challenge driving detector design at \mumu colliders is the reconstruction of a broad range of collision products-of-interest 
in the presence of beam-induced background (BIB) arising from muon decay along the beamline~\cite{Lucchesi:2020dku}. 
BIB consists primarily of TeV-scale electrons that interact with the accelerator and detector components. This process generates large multiplicities of predominantly soft secondary particles, some of which enter the detector volume. These secondary particles are typically out-of-time with respect to the beam collisions and do not originate from the interaction region. As a result, the presence of BIB imposes stringent demands on the granularity, resolution, and timing capabilities of muon collider detectors to effectively suppress background to acceptable levels without significantly degrading experimental performance. Additionally, radiation tolerance is crucial, as detectors must withstand moderate radiation doses induced by the beam.

Past studies of muon collider detectors, including those the context of previous R\&D efforts such as the
US Muon Accelerator Program (MAP), have mainly focused on lower center-of-mass
energies~\cite{Ankenbrandt:1999cta,Boscolo:2018ytm,Neuffer:IPAC2017-TUPIK038}. More recently, comprehensive
simulations of a detector designed for a 3~TeV muon collider, adapted from a 3~TeV CLIC detector concept, have demonstrated
robust performance~\cite{accettura2023muoncollider}. However, this design is 
inadequate for 10~TeV collisions due to significant differences in final state particle kinematics,
such as energy, rapidity, and decay length. Furthermore, the particle spectrum and timing characteristics of 
the BIB are notably different between the two energy scales, as the machine-detector interface and collider
lattice have been reoptimized for 10~TeV~\cite{InternationalMuonCollider:2025sys}. It is therefore
critical to develop new detector
designs for 10~TeV leptonic collisions to serve as the basis for physics studies, leverage anticipated 
advancements in detector technologies, and maximize the physics potential of a collider at this energy 
scale. Key physics goals include precision tests of the Higgs boson and other SM parameters as well as 
excellent sensitivity to a broad range of exotic phenomena beyond the SM, underscoring the importance
of developing a robust, general-purpose detector.

Two initial concepts that build on the 3~TeV detector have recently emerged that make different design 
choices and explore the use of different technologies: MAIA (Muon Accelerator Instrumented Apparatus)
and MUSIC (MUon System for Interesting Collisions)~\cite{Andreetto:2025mrd}. Both designs represent a step
forward in the evolution of lepton collider detectors towards the higher energy regime of 3~TeV and beyond. 
This paper describes MAIA in detail and evaluates its current simulated performance.

The structure of the paper is as follows: Section~\ref{sec:exp_conditions} provides an overview of the experimental conditions and the modelling of the BIB. The detector layout is described in Section~\ref{sec:detector_layout}. Section~\ref{sec:simulation} details the detector simulation setup, while the simulated performance results are presented in Section~\ref{sec:performance}. Future work and potential improvements are outlined in Section~\ref{sec:future}, with a brief summary provided in Section~\ref{sec:conclusions}.

\section{Experimental conditions}
\label{sec:exp_conditions}

The muon decays occurring in the proximity of the interaction region contribute a significant source of BIB, which play a major role in driving detector design. In the context of this study, we only consider decay-induced background and incoherent pair production~\cite{ginzburg1996ee, Schulte:382453} occurring at the interaction point in the detector reconstruction. 
Other sources of background, such as the beam halo losses~\cite{halo-calzolari} in the proximity of the interaction region, are not considered for this study.

The collider lattice under consideration for this study is the EU24 (v0.8) design developed by the IMCC~\cite{skoufaris:ipac22-mopotk031, skoufaris:ipac23-mopl064}. This design is based on a conventional quadrupole triplet scheme for the final focusing, where the distance from the interaction point to the first quadrupole is \mbox{$L^{*}=\SI{6}{m}$}.
Upstream of the final focus triplet, a dipolar chicane is included in the lattice design to suppress the contribution of muon decays occurring in the long straight section between the triplet and the chromaticity correction section. Secondary electrons and positrons produced in these distant decays are deflected more strongly than the primary beam and rapidly lose energy in the dipole field, causing them to be swept into the machine aperture before reaching the detector region~\cite{calzolari2023jacow}.
Muon decays in the final focusing region account for the vast majority of the BIB. In contrast, decays within the dipole elements produce secondary particles that are rapidly deflected into the beam aperture, rendering their contribution to detector backgrounds negligible. The nominal beam intensity assumed for simulations is $\SI{1.8E12}{}$ muons per bunch, with a transverse normalized emittance of $\SI{25}{\micro m\, rad}$.

\begin{figure}[!b]
    \centering
    \includegraphics[width=0.6\linewidth] {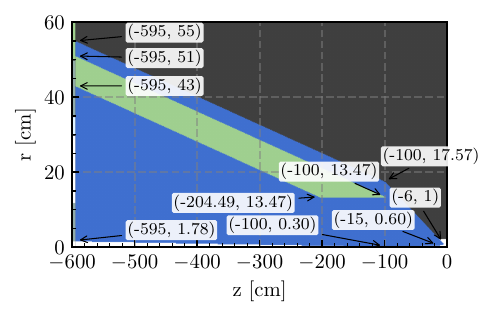}
    \caption{Geometrical description of a single nozzle, with measurements reported in cm. Both nozzles are cylindrically symmetric symmetry about the z-axis.  The green component is borated polyethylene, while the blue-gray area represents tungsten heavy alloy (W-alloy). The area in black is dedicated to the detector elements, which are not shown.}
    \label{fig:nozzle_geo}
\end{figure}

To mitigate the background coming from these secondary particles, past studies suggested placing dedicated shielding equipment between the beamline and the detector. In particular, in the context of MAP, a nozzle-like shape optimized for collider energies up to $\SI{1.5}{TeV}$ was proposed \cite{mokhov2012muon, Curatolo:2804499}. That conceptual nozzle design contains two main elements surrounding the beryllium beam pipe: an inner tungsten heavy alloy layer to mitigate the electromagnetic showers and a borated polyethylene cladding to moderate and capture the neutronic component. The nozzle implemented in this study, illustrated in Figure~\ref{fig:nozzle_geo}, is based on the MAP design and was adapted for a $\SI{10}{TeV}$ collider environment~\cite{InternationalMuonCollider:2025sys}. The nozzle design plays a significant role in background mitigation, and its configuration must be jointly optimized with the collider lattice and detector design~\cite{calzolari2022sissa, calzolari2023jacow,Calzolari:2025mcq}.

The BIB is simulated with FLUKA~\cite{BATTISTONI201510, FLUKA-new}. The simulation begins with beam-muon decays sampled from the matched beam phase space. The electrons and positrons produced in the decays are propagated along the collider until they reach the detector volume, at which point the secondary particle propagation is stopped and the particle type, position, energy-momentum, and timing information is stored. 

Neutrons are simulated down to thermal energies, and all other particle species are simulated down to $\SI{100}{keV}$ kinetic energy. The latter cut is chosen since, below this threshold, the probability of producing a hit in the vertex detector is negligible. Beyond the default FLUKA physics settings, additional phenomena are enabled, such as synchrotron radiation, photo-nuclear interactions, and electro-nuclear interactions, to enhance simulation accuracy.

\begin{figure}[!ht]
 \centering
 \subfloat[\label{fig:bib10_energy}]{\includegraphics[width=0.49\textwidth]{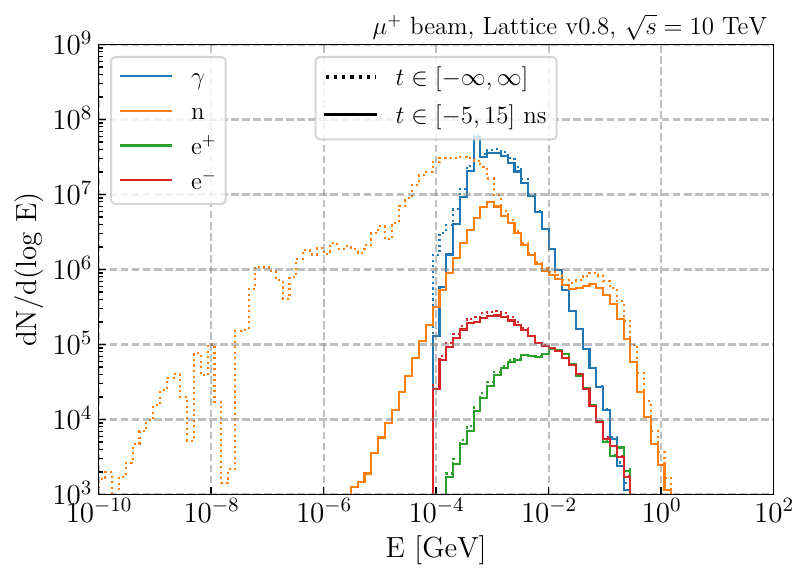}}
 \subfloat[\label{fig:bib10_time}]{\includegraphics[width=0.49\textwidth]{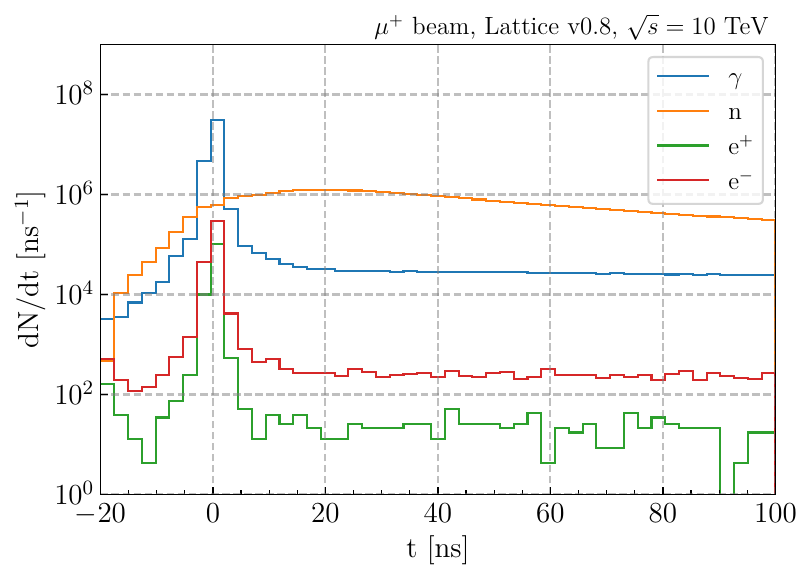}} \\
 \subfloat[\label{fig:bib10_zdist}]{\includegraphics[width=0.49\textwidth]{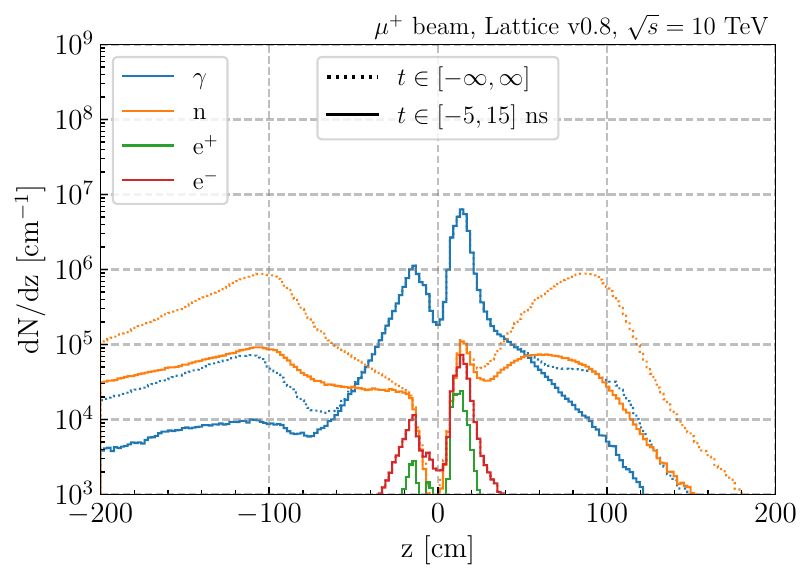}}
 \subfloat[\label{fig:bib10_zmu}]{\includegraphics[width=0.49\textwidth]{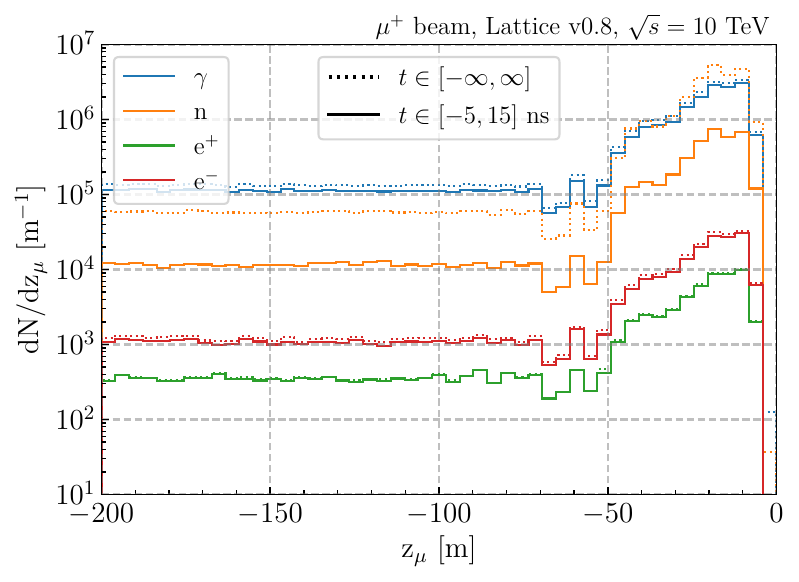}}
  \caption{Various spectra of BIB particles as generated by FLUKA. All the results are normalized to the BIB generated during a single bunch crossing at a nominal bunch intensity of $\SI{1.8E12}{}$ muons per bunch. Only the main particle components are reported (i.e., electrons, positrons, photons, and neutrons). At the top, the \protect\subref{fig:bib10_energy} energy and \protect\subref{fig:bib10_time} time distribution spectra are shown. At the bottom, \protect\subref{fig:bib10_zdist} the distribution of the particle z position at the exit of the nozzle elements, and
  \protect\subref{fig:bib10_zmu} the total number of secondary particles as a function  of the longitudinal muon decay position are shown. Muon decays occurring outside of the final focusing region contribute negligibly to the BIB.}
 \label{fig:BIB_plot}
\end{figure}

\begin{figure}[!ht]
\includegraphics[width=0.9\textwidth]{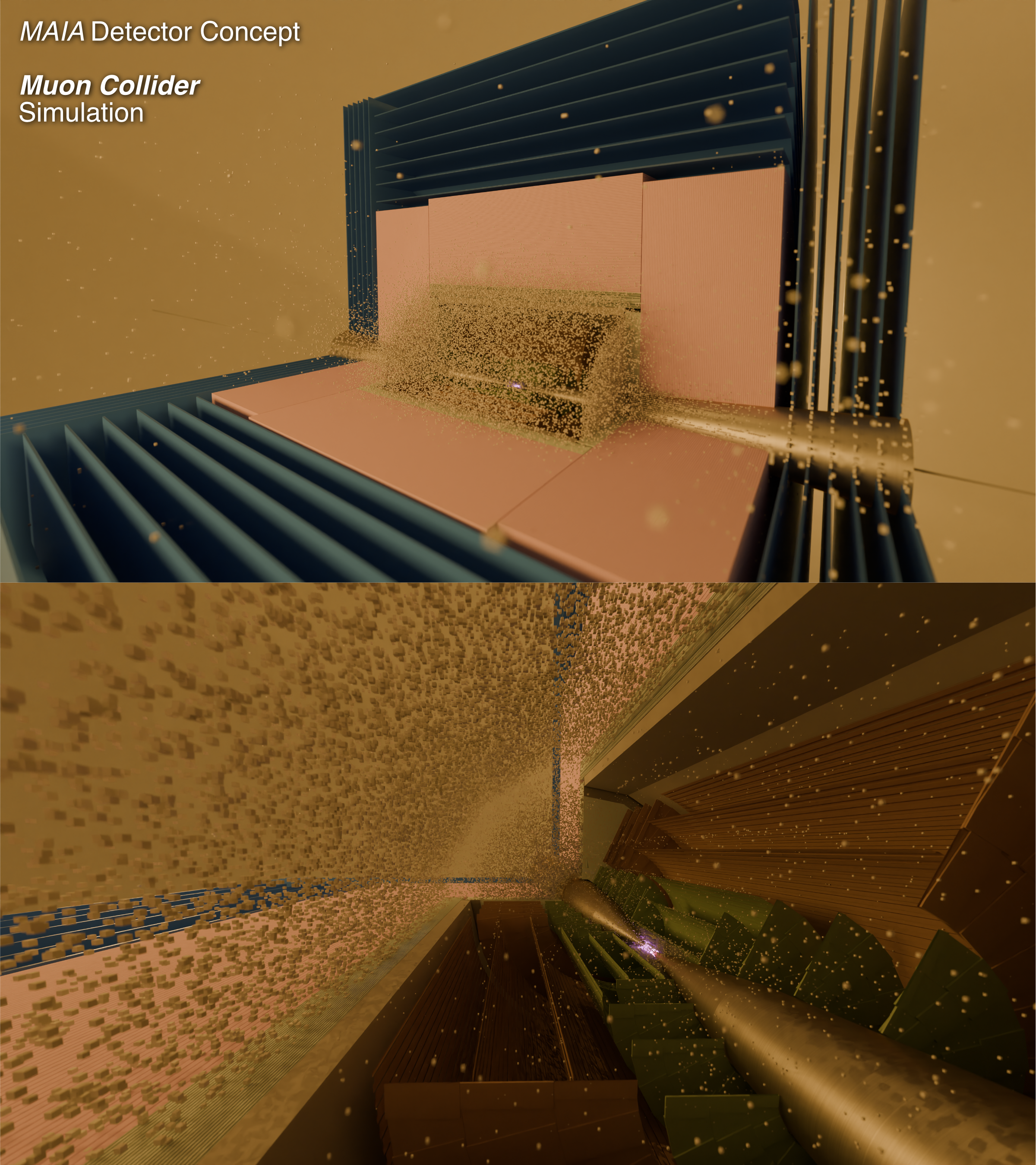}
 \caption{Illustration of a simulated BIB event in the MAIA detector with a cutaway of $\pi/2$ in $\phi$. BIB energy depositions in the calorimeters are shown as boxes, and the largest contributions are located in the ECAL and the Vertex Detector.}
 \label{fig:bib_maia}
\end{figure}

The various BIB particle spectra are reported in Figure~\ref{fig:BIB_plot}. The energy spectra exemplify the effectiveness of the nozzle boron layer since the thermal neutron component is significantly suppressed. Considering the photon component, the peak in the spectrum contains both the characteristic gamma line in the boron neutron capture ($\SI{477}{keV}$) and the annihilation peak ($\SI{511}{keV}$). An event display illustrating the impact of the simulated BIB in the detector is shown in Figure ~\ref{fig:bib_maia}.

The position where the secondary BIB particles leave the nozzle to enter the detector region plays a dominant role in the impact of BIB on the detector, and thus, detector design requirements. For example, BIB particles departing from the nozzle in proximity to the tip directly impact the innermost tracker layer, leading to increased occupancies.

The generation of the incoherent pairs was performed using the GUINEA-PIG code~\cite{Schulte:1997nga}, a Monte Carlo event generator developed to simulate beam–beam interactions in lepton colliders. 
The GUINEA-PIG program calculates the full electromagnetic interaction between the two bunches,
producing a list of incoherent $e^{+}e^{-}$ pairs at the interaction point. 
The generated pairs are subsequently injected into the FLUKA simulation as primary particles and propagated beyond the beampipe into the detector region. The FLUKA run corresponds to a full bunch crossing at nominal beam intensity \cite{calzolari_2025_17494967}.

\section{Detector layout}
\label{sec:detector_layout}

The conceptual detector layout is conceived for the highest possible hermiticity given
the constraints of the shielding nozzles described above. The detector is designed to be approximately azimuthally symmetric with varying $n$-fold symmetries across the subdetectors, and full azimuthal coverage is assumed. The BIB mitigation is the primary challenge to approach $4\pi$ detector solid-angle coverage. The nozzle extends down to pseudorapidities as low as $|\eta|=2.44$, corresponding to approximately $10^{\circ}$ in polar angle $\theta$ with respect to the beam axis. While the detector can therefore cover the region $|\eta|<2.44$ without interference with the nozzles, regions of the detector closer to the nozzles experience additional background due to BIB. 

At the high muon collision energy, the detector must be designed to ensure high-resolution measurements of multi-TeV physics objects. This requires large tracking volumes and a high solenoid field to to resolve high-$\pT{}$ charged particles and deep calorimeters minimize punch-through of energetic objects. The spatial dimensions and primary materials of each subdetector are described in Table~\ref{tab:detector-dimensions}.

\begin{table}[!ht]
\footnotesize
    \centering
    \begin{tabular}{r|l||c|c|l}
        \textbf{Subsystem} & \textbf{Region} & \textbf{R dimensions [cm]} & \textbf{$|$Z$|$ dimensions [cm]} & \textbf{Material} \\
        \hline\hline
        Vertex Detector & Barrel & $3.0 - 10.4$     & $65.0$            & Si \\
                        & Endcap & $2.5 - 11.2$     & $8.0 - 28.2$      & Si \\
        \hline
        Inner Tracker   & Barrel & $12.7 - 55.4$    & $48.2 - 69.2$     & Si \\
                        & Endcap & $40.5 - 55.5$    & $52.4 - 219.0$    & Si \\
        \hline
        Outer Tracker   & Barrel & $81.9 - 148.6$   & $124.9$           & Si \\
                        & Endcap & $61.8 - 143.0$   & $131.0 - 219.0$   & Si \\
        \hline\hline
        Solenoid        & Barrel & $150.0 - 185.7$  & $230.7$           & Al \\
        \hline\hline
        ECAL            & Barrel & $185.7 - 212.5$  & $230.7$           & W + Si \\
                        & Endcap & $31.0 - 212.5$   & $230.7 - 257.5$   & W + Si \\
        \hline
        HCAL            & Barrel & $212.5 - 411.3$  & $257.5$           & Fe + PS \\
                        & Endcap & $30.7 - 411.3$  & $257.5 - 460.0$   & Fe + PS \\
        \hline\hline
        Muon Detector   & Barrel & $415.0 - 589.5$  & $460.0$           & Air + RPC \\
                        & Endcap & $44.6 - 589.5$   & $460.0 - 620.0$   & Air + RPC \\
    \end{tabular}
    \caption{Boundaries and materials of individual subdetectors.}
    \label{tab:detector-dimensions}
\end{table}

\begin{figure}
    \centering
    \includegraphics[width=0.95\linewidth]{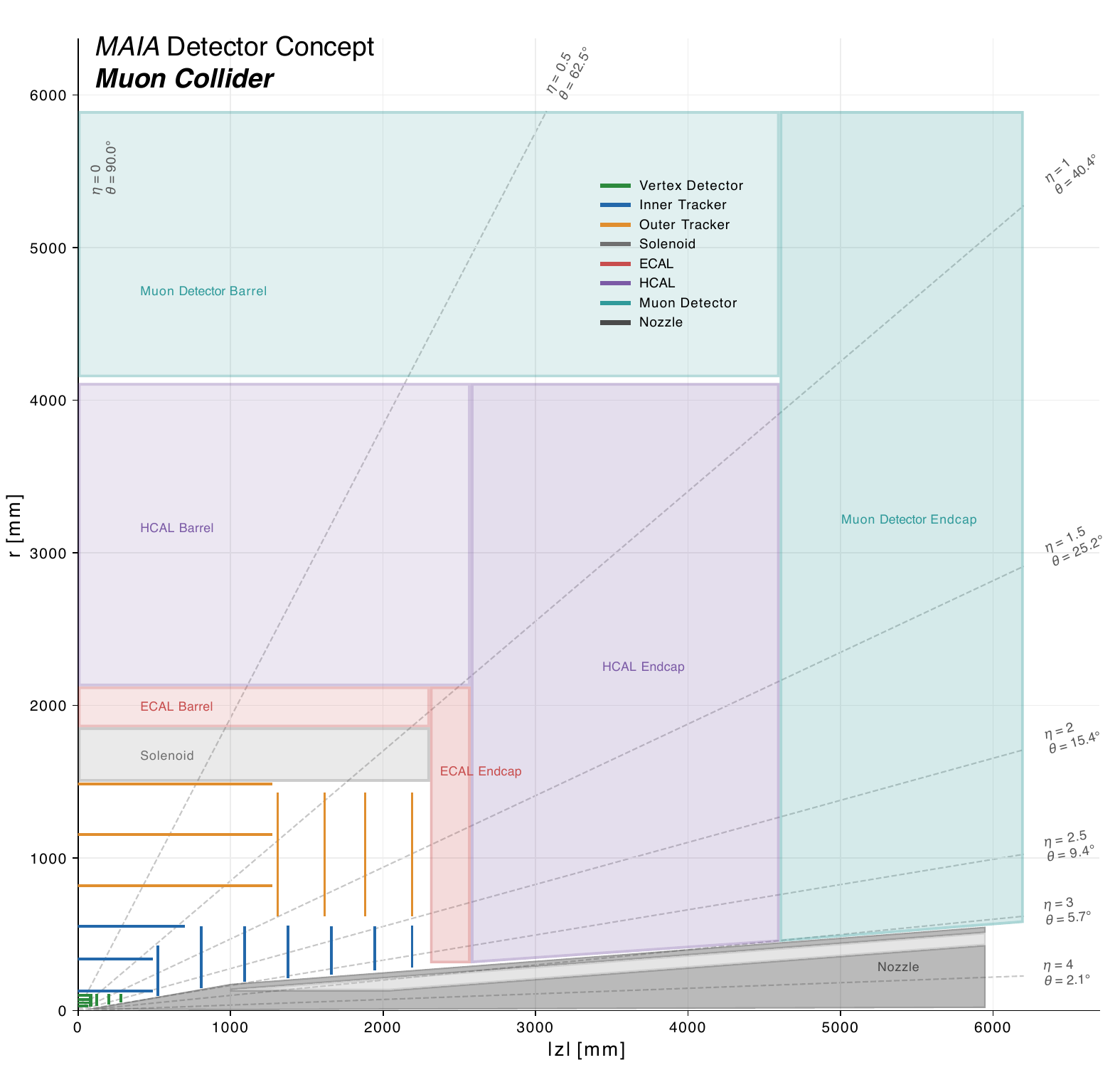}
    \caption{An $r-z$ schematic of one quarter of the MAIA detector. In addition to the identified subsystems, the nozzle is shown in gray, with a darker shade corresponding to tungsten and a lighter shade corresponding to a borated polyethylene layer.}
    \label{fig:rzdet}
\end{figure}

\begin{figure}[!b]
\includegraphics[width=0.8\textwidth]{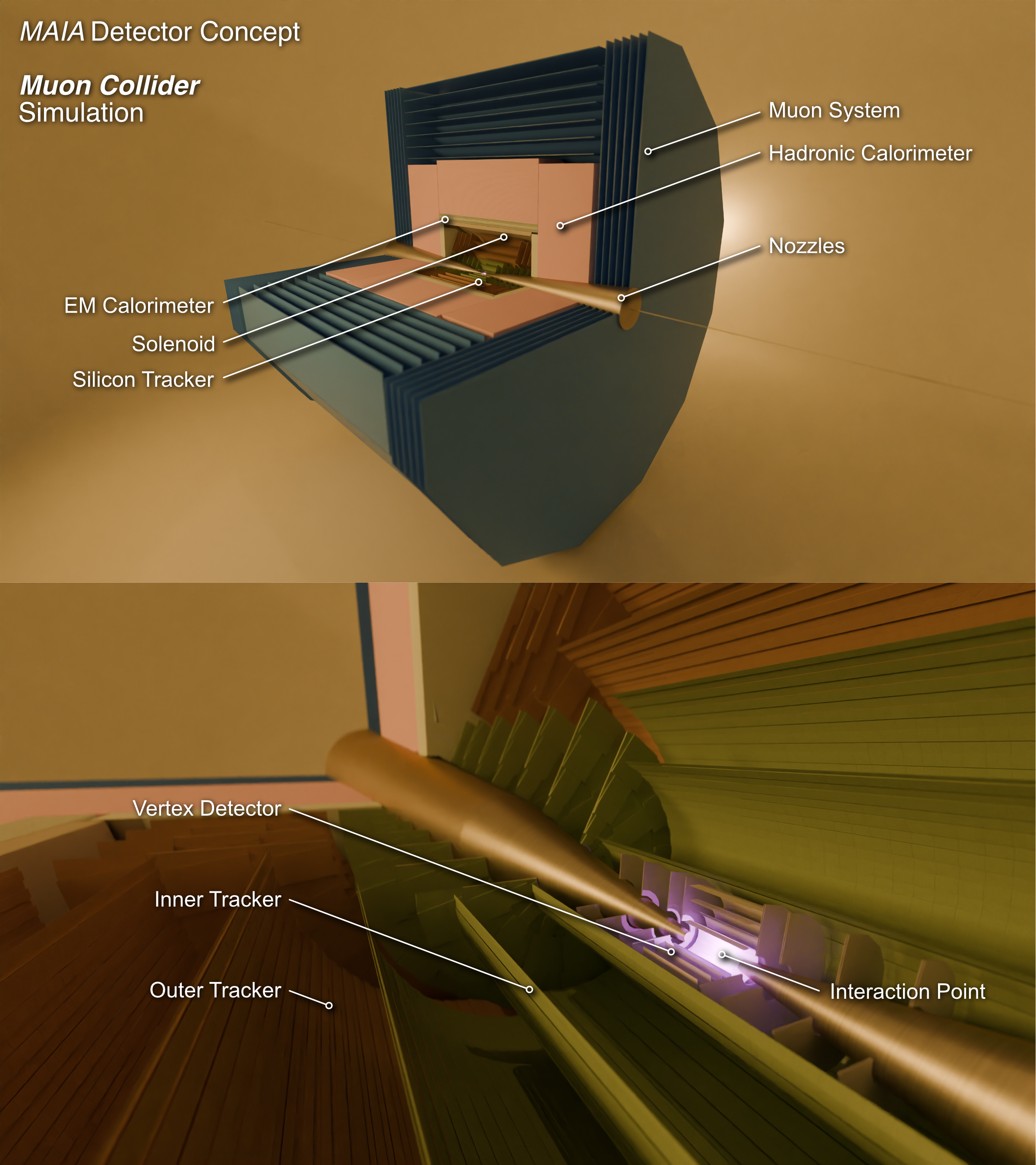}
\caption{Illustration of the MAIA detector layout, including a closeup of the silicon tracker around the interaction point. The detector is shown with a $\pi/2$ cutaway in $\phi$ for illustration.}
\label{fig:subdetectors}
\end{figure}

An overview of the detector subsystem layout is given in Figure~\ref{fig:subdetectors} and a schematic diagram is shown in Figure~\ref{fig:rzdet}. The conceptual design of each subdetector will be discussed in the following sections.

\FloatBarrier
\subsection{Tracking detectors}

\begin{figure}
    \centering
    \includegraphics[width=0.95\linewidth]{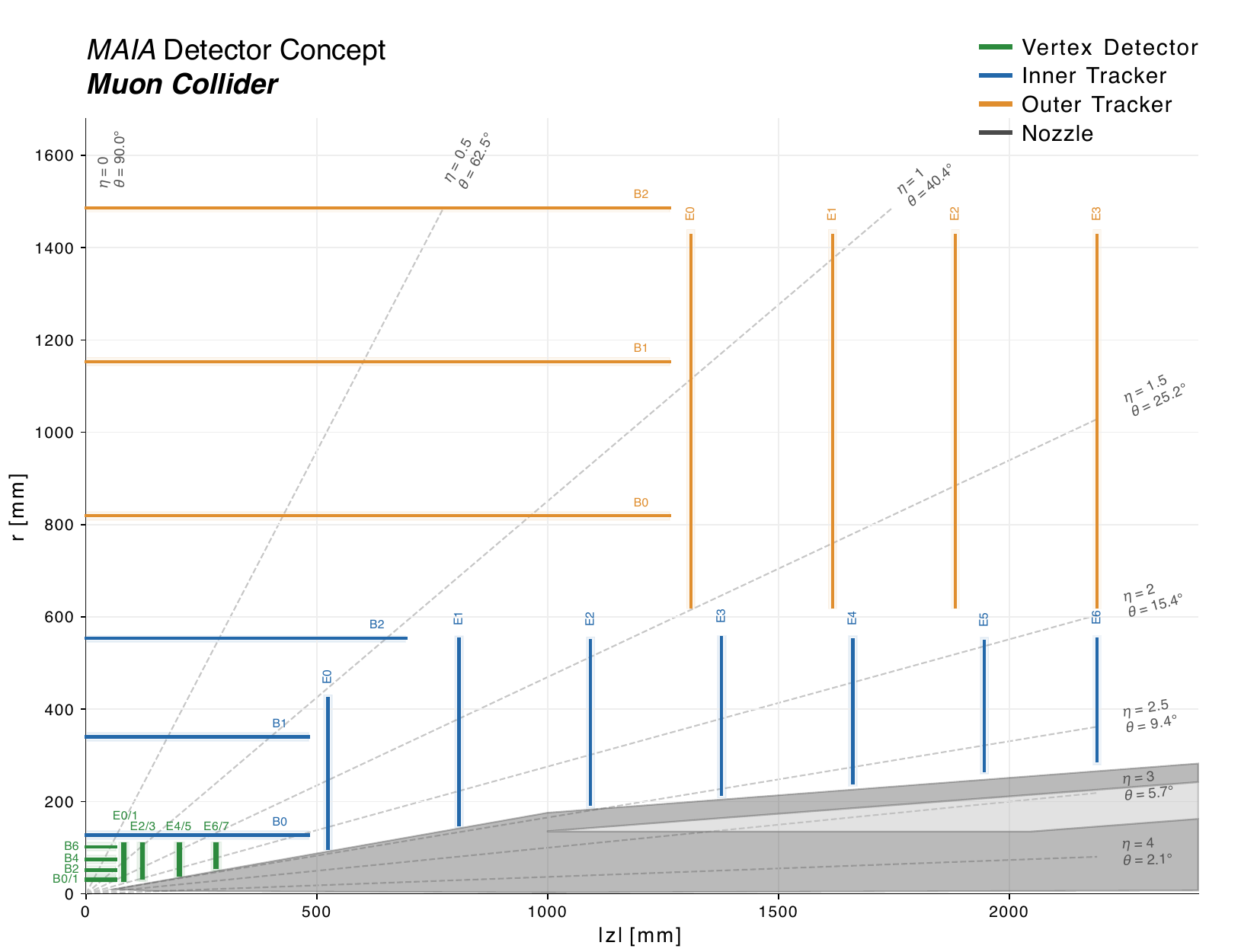}
    \caption{An $r-z$ schematic of one quarter of the MAIA tracker showing the Vertex Detector (green), the Inner Tracker (blue), and the Outer Tracker (orange). Double-layers in the Vertex Detector are indicated with paired indices. The nozzle is shown in gray, with a darker shade corresponding to tungsten and a lighter shade corresponding to a borated polyethylene layer.}
    \label{fig:rztrack}
\end{figure}

The conceptual tracker design comprises three silicon subdetectors summarized in Table~\ref{tab:tracker_specs}: the Vertex Detector, a high-resolution pixel detector closest to the collision point; the Inner Tracker, a macro-pixel detector; and the Outer Tracker, a second macro-pixel detector with larger cells and reduced granularity. The layout and specifications of these three sub-systems are based on those of the $\sqrt{s}=3$~TeV detector design~\cite{Accettura:2023ked}, with some modifications to account for both higher collision energies and a higher solenoid magnetic field. A schematic diagram can be found in Figure~\ref{fig:rztrack}.

\begin{table}[b!]
    \centering
    \begin{tabular}{l|c|c|c}
                        &  \textbf{Vertex Detector} & \textbf{Inner Tracker} & \textbf{Outer Tracker} \\
    \hline\hline
    Sensor type           &  pixels & macro-pixels & macro-pixels \\
    Barrel Layers  &  4  & 3  & 3 \\
    Endcap Layers (per side)  &  4   &  7 & 4 \\
    Cell Size           &  \qty{25}{\um} $\times$ \qty{25}{\um} & \qty{50}{\um} $\times$ \qty{1}{\mm} & \qty{50}{\um} $\times$ \qty{10}{\mm} \\
    Sensor Thickness    &  \qty{50}{\um} & \qty{100}{\um} & \qty{100}{\um} \\
    Time Resolution     &  \qty{30}{\ps} & \qty{60}{\ps} & \qty{60}{\ps} \\
    Spatial Resolution  &  \qty{5}{\um} $\times$ \qty{5}{\um} & \qty{7}{\um} $\times$ \qty{90}{\um} & \qty{7}{\um} $\times$ \qty{90}{\um} \\
    \end{tabular}
    \caption{Spatial and time resolution assumptions for Tracking Detector sub-systems. There is no resolution difference between the barrel and end-cap regions. The first layer of the Vertex barrel and all Vertex endcap layers are implemented as double layers.}
    \label{tab:tracker_specs}
\end{table}

The occupancy and time structure of BIB particles motivate a highly granular detector with precision timing in every layer. In each bunch crossing, BIB particles result in hit densities of around $30000~\mathrm{hits/cm}^{2}$ in the innermost pixel layer. 
Pixels with $25\times\SI{25}{\micro\meter}^2$ granularity and $\SI{30}{\pico\second}$ resolution timestamps are required to achieve a desired occupancy of $1\%$. In comparison, the ATLAS and CMS experiments will implement timing detectors with ${\sim}1\times\SI{1}{\milli\meter}^2$ pixels and $\SI{30}{\pico\second}$ resolution based on Low Gain Avalanche Detectors in the outermost layer of the tracker~\cite{HGTD:2091129,Butler:2019rpu}. While there are are several promising sensor technologies for muon collider detector needs, the smaller pixel size poses the largest challenge for front-end power consumption and readout~\cite{MuonCollider:2022glg}. The detector must also survive an expected $1~\mathrm{MeV}$ neutron equivalent fluence of $1\times10^{14-15}~\mathrm{neq/cm}^2$ and a total ionizing dose range of $10^{2-5}~\mathrm{Gy}$ per year.

\begin{figure}[b!]
 \centering
 \includegraphics[width=0.72\textwidth]{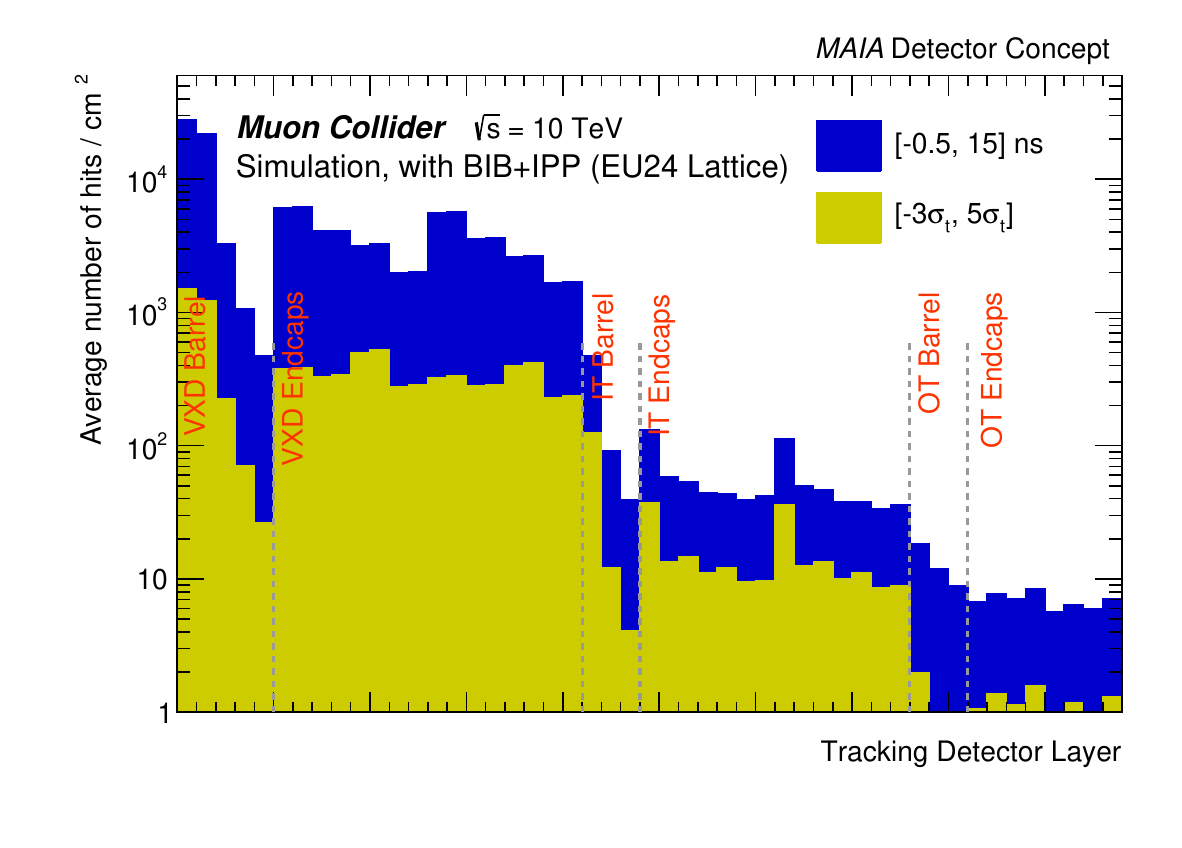}
 \caption{Average hit density in the tracking detectors, shown separately for each detector layer, for $\sqrt{s}=10$~TeV BIB using the EU24 collider lattice. The yellow histograms include a selection on the hit time, corrected by the time of arrival for particles traveling at the speed of light. The selection applies a window of $[-3 \sigma_t, 5 \sigma_t]$ from the beam crossing, where $\sigma_t$ refers to the detector time resolution.
The dashed lines highlight the boundaries between sub-detectors and the orange text specifies the sub-detector name. The bins are ordered respectively by ascending radius for barrel detector and ascending $|z|$ for endcap detectors, with $z<0$ layers preceding those with $z>0$.}
 \label{fig:tracker_occupancy}
\end{figure}

\begin{figure}[t!]
 \centering
 \subfloat[\label{fig:tracker_occupancy_z}]{\includegraphics[width=0.49\textwidth]{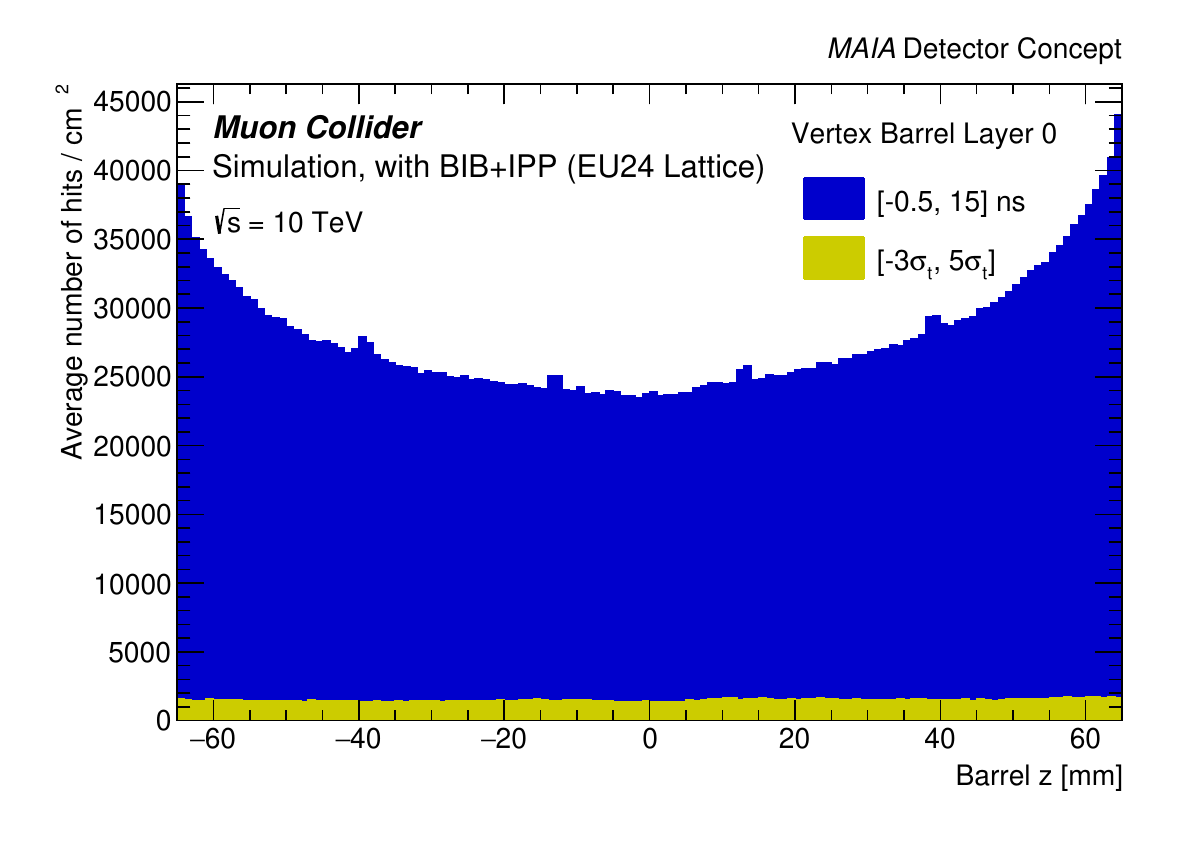}}
 \subfloat[\label{fig:tracker_occupancy_R}]{\includegraphics[width=0.49\textwidth]{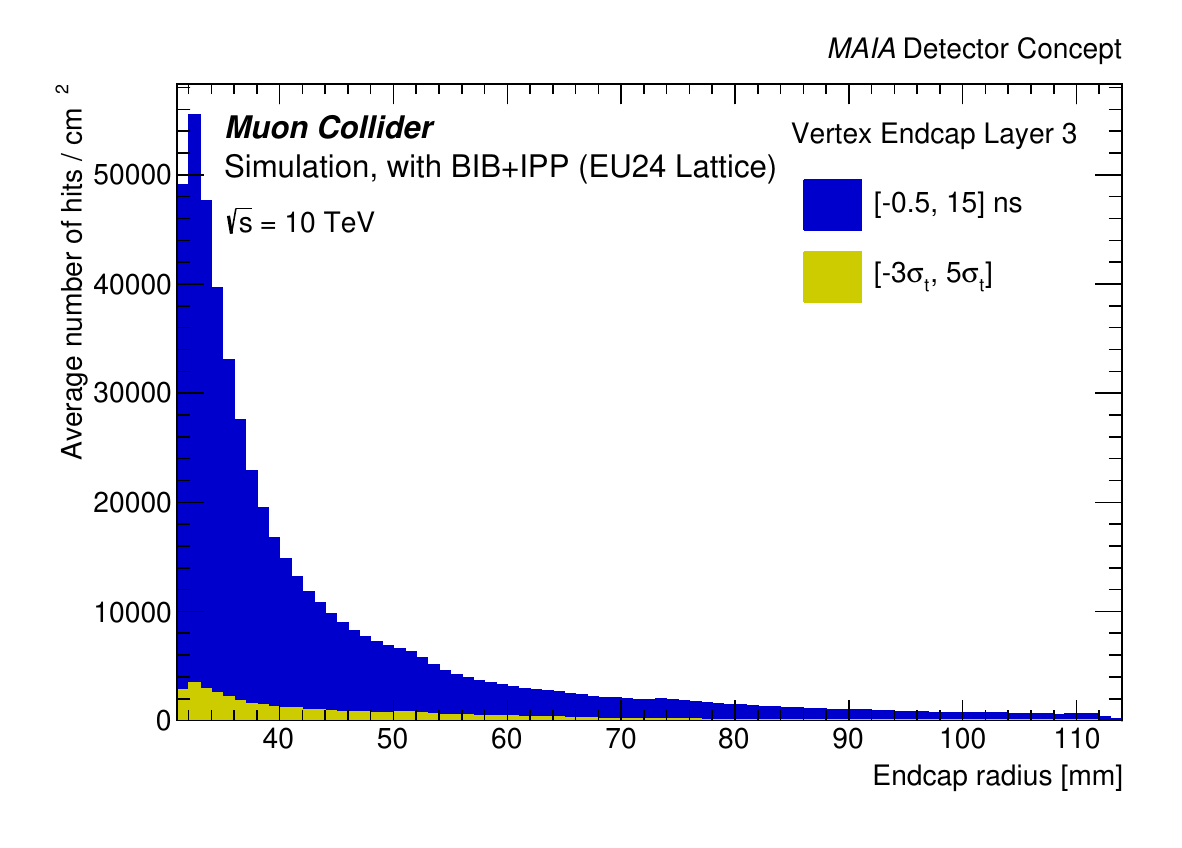}}
 \caption{Hit density in the highest average occupancy layers in the vertex detector for \protect\subref{fig:tracker_occupancy_z} the barrel and \protect\subref{fig:tracker_occupancy_R} the endcap. Densities are shown as a function of the longitudinal or radial position, for $\sqrt{s}=10$~TeV BIB.}
 \label{fig:tracker_occupancy_detail}
\end{figure}

The BIB hit density decreases rapidly with radial distance from the beamline, as shown in Figure~\ref{fig:tracker_occupancy}.
BIB hits are not uniformly distributed on each tracker plane. Figure~\ref{fig:tracker_occupancy_detail} illustrates the main dependencies for layers with the highest average occupancy of the vertex detector. In both figures, stringent timing cuts of $[-3\sigma_t, 5\sigma_t]$ significantly reduce the BIB occupancy, demonstrating the importance of precise timing capabilities. This selection also almost entirely removes the $z$-dependence in the barrel region, with limited mitigation on the radial dependence in the endcap disks.

One notable difference with respect to the $3$~TeV design is the reduction in doublet layers in the Vertex Detector. Doublet silicon layers, similar to those used in the High-Luminosity LHC (HL-LHC) upgrade to the CMS tracker~\cite{CERN-LHCC-2017-009}, are used to reduce the computational complexity of track reconstruction. 
For the $3$~TeV design, all Vertex Detector layers were implemented as doublet layers.
The $3$~TeV design also used a Conformal Tracking algorithm, optimized for the clean environment of $e^+e^-$ colliders~\cite{Brondolin:2019awm}, which required doublet layers to reduce the time needed to reconstruct an event. As discussed below in Section~\ref{sec:performance}, an algorithm developed for the busier environment of hadron colliders is now used for track reconstruction.
Doublet layers in the outer three layers of the Vertex Detector are no longer necessary due to this improvement, so they were removed.

\subsection{Magnet system}
\label{subsec:magnet}

The tracking system for the MAIA detector concept is immersed in a $5~\mathrm{T}$ solenoidal magnetic field, in contrast to the $3.57~\mathrm{T}$ solenoid of the 3~TeV detector design. The increased magnetic field maintains precision momentum resolution for the highest momentum charged particles produced in collisions. The higher field also reduces the radius of curvature for low momentum particles, resulting in reduced occupancy with respect to $\sqrt{s}=3~\mathrm{TeV}$. To accommodate the increased field strength and the engineering challenges associated with producing a large-aperture magnet, the 10~TeV detector design places the solenoid inside the calorimeter system, in contrast to the 3~TeV design, which locates the solenoid outside the calorimeters. In the current MAIA design, there is no magnetic field in the muon system.

\subsection{Calorimetry}
\label{subsec:calorimetry}

The MAIA detector concept makes use of a silicon-tungsten electromagnetic calorimeter (ECAL) and an iron-scintillator hadronic calorimeter (HCAL). Both systems are based on the CLIC calorimeter design~\cite{robson2018compactlineareecollider}, which was also the starting point for earlier iterations of a muon collider detector optimized for $\sqrt{s}=3$~TeV~\cite{Accettura:2023ked}. Compared to the calorimeters optimized for lower energy, this detector concept has more layers, and each layer has a slightly thicker absorber. The cells have also been slightly scaled up in size. More details on the parameters of the calorimeter can be found in Table~\ref{tab:calo_specs}. 

\begin{table}[b!]
    \centering
    \begin{tabular}{l|c|c}
                        &  \textbf{Electromagnetic Calorimeter} & \textbf{Hadron Calorimeter} \\
    \hline\hline
    Cell type           &  Silicon - Tungsten & Iron - Scintillator \\
    Cell Size           &  \qty{5.1}{\mm} $\times$ \qty{5.1}{\mm} & \qty{30.0}{\mm} $\times$ \qty{30.0}{\mm}  \\
    Sensor Thickness    &  \qty{0.5}{\mm} & \qty{3.0}{\mm}  \\
    Absorber Thickness  &  \qty{2.2}{\mm} & \qty{20.0}{\mm}  \\
    Number of layers  &  50 & 75  \\
    \end{tabular}
    \caption{Cell and absorber sizes in the calorimeter systems, describing both the barrel and end-cap regions.}
    \label{tab:calo_specs}
\end{table}

The design of the calorimetry for this detector is a modest extension of current technology. The technical specifications of the ECAL's silicon are similar to that of the CMS high granularity calorimeter (HGCAL), currently under construction for the HL-LHC upgrade~\cite{CERN-LHCC-2017-023}. The cell sizes are slightly smaller in the MAIA design, and the sensors are slightly thicker, but the overall parameters are similar.
The HGCAL will have a 50 ps timing resolution. The MAIA HCAL is similar in design to the ATLAS Tile Calorimeter, albeit with smaller cells by roughly a factor of 10 in each direction \cite{ATLAS:1996aa}, and a timing resolution of roughly 1 ns. MAIA's ECAL and HCAL simulations have not yet assumed similar timing resolution to their LHC equivalents, but instead currently sum all deposits made from -300 to 300 ps with respect to the expected time of arrival. 

\begin{figure}[!ht]
 \centering
 \includegraphics[width=0.7\textwidth]{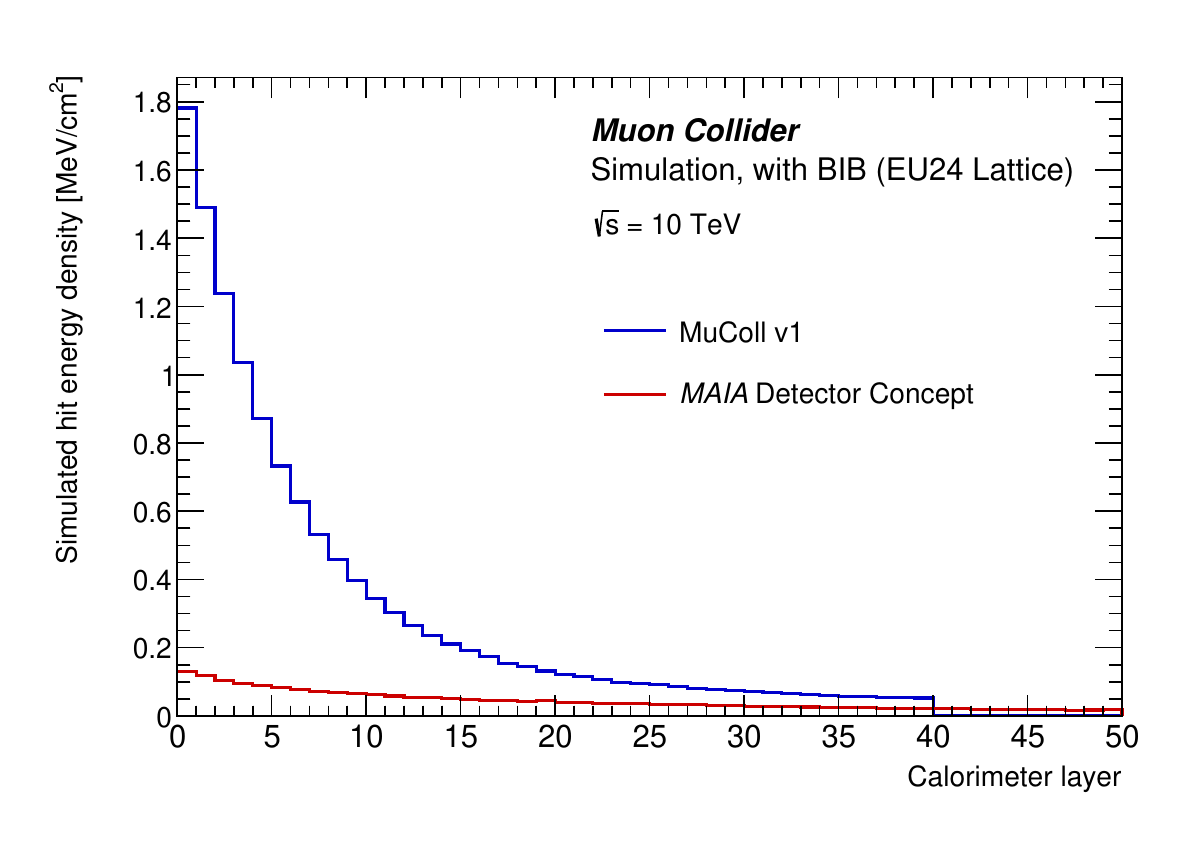}
 \caption{Average energy density of simulated BIB hits for each ECAL barrel layer. The layer index increases with radius. The same BIB particles are propagated through the 3~TeV detector (MuColl v1) and the MAIA concept to illustrate the effects of the solenoid placing.}
 \label{fig:ECAL_energy_density}
\end{figure}

A significant change in the MAIA detector relative to the 3~TeV detector is the relocation of the solenoid to inside the calorimeters, as discussed. The solenoid adds approximately 265~mm of aluminum and thinner steel layers in the region $|z|\lesssim2.3$~m, corresponding to around 4~$X_{0}$ and 1~$\lambda$ for a particle crossing the material in the transverse direction. This material reduces the incoming BIB flux by around an order of magnitude in the central region of the detector: Figure~\ref{fig:ECAL_energy_density} shows the energy density per layer comparing the proposed 10~TeV layout with the 3~TeV detector design. Of course, this shielding also impacts the signal, which is addressed in a dedicated calibration procedure discussed in Section~\ref{sec:performance}.

Compared to the tracker, the calorimeter faces distinct challenges for BIB resilience: it has a higher radius, resulting in a lower flux, and its energy-proportional response means that high-energy signatures can be picked out with relative ease. However, the larger cell sizes and longer integration times associated with most calorimeter technologies mean that BIB still poses challenges to calorimetry.

\begin{figure}[b!]
 \centering
 \subfloat[\label{fig:ECAL_BIB_energy_L0}]{\includegraphics[width=0.5\textwidth]{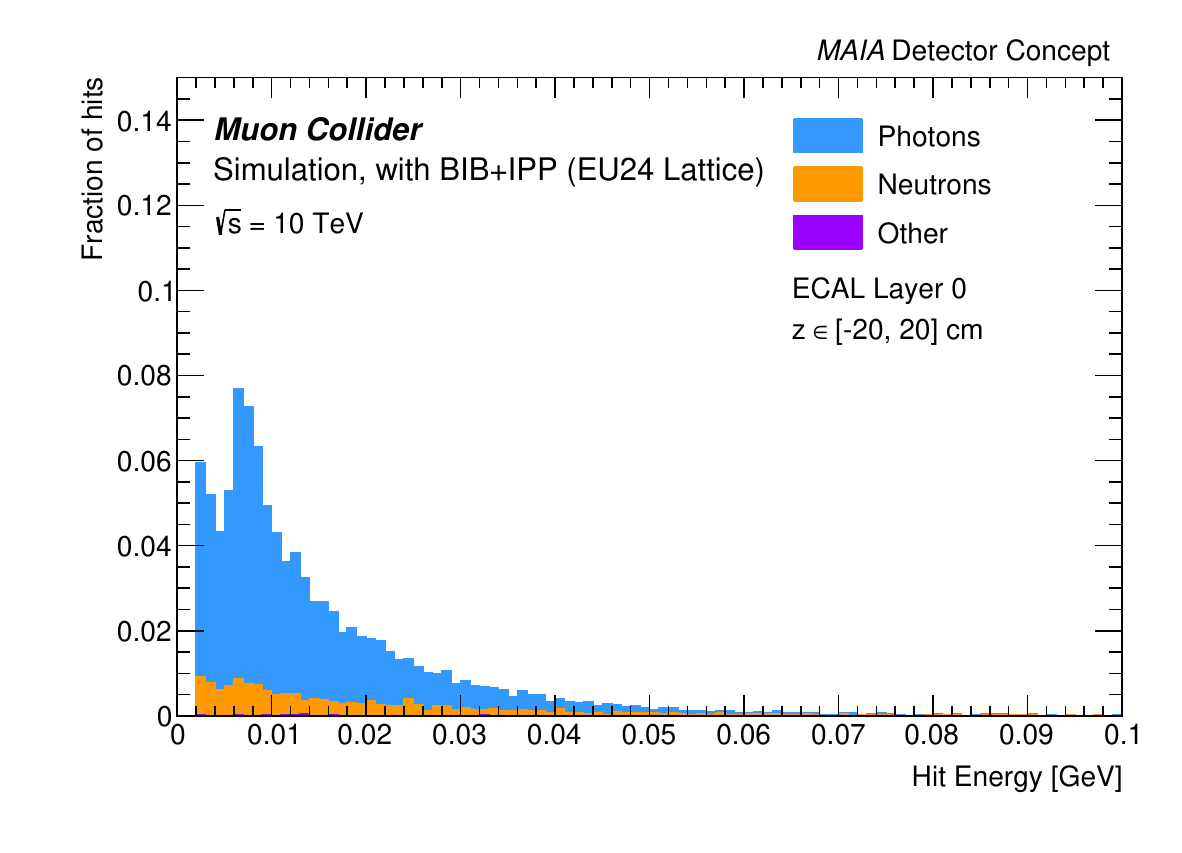}}
 \subfloat[\label{fig:ECAL_BIB_energy_L15}]{\includegraphics[width=0.5\textwidth]{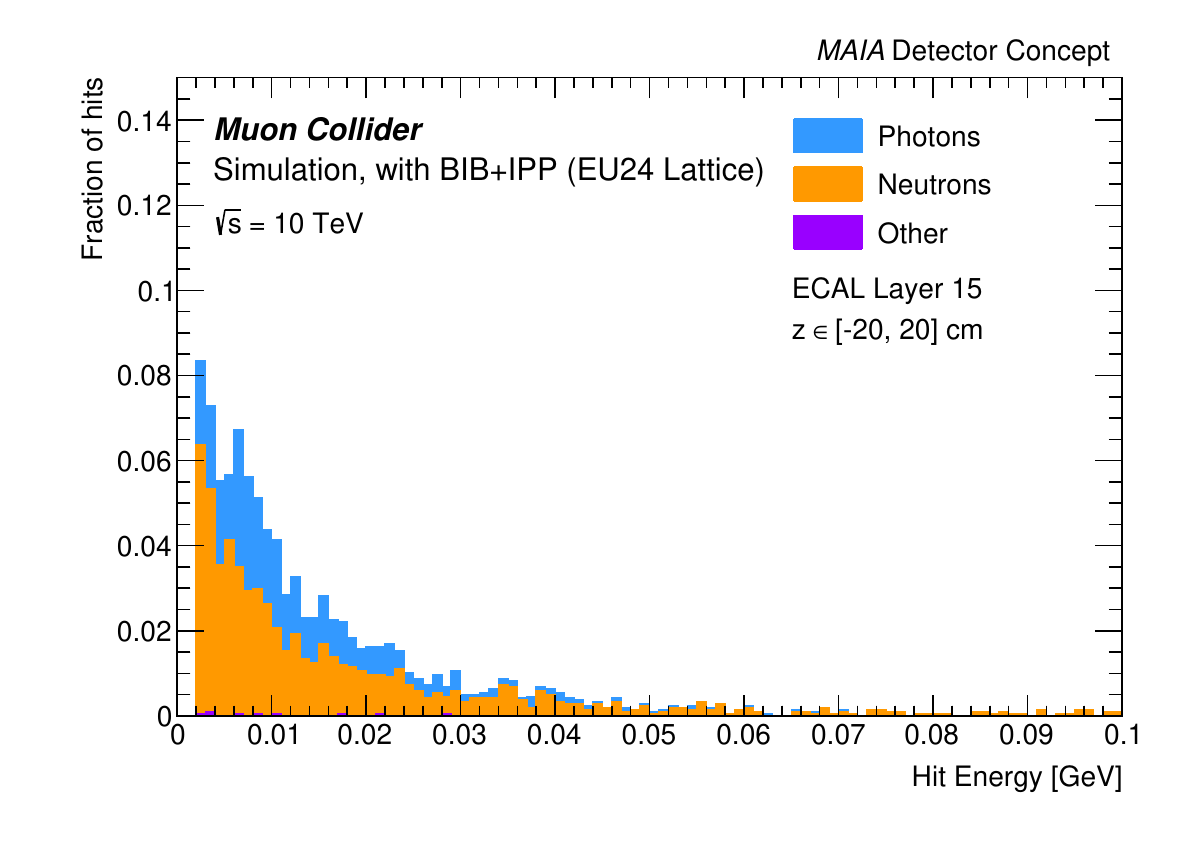}} \\
 \subfloat[\label{fig:ECAL_BIB_energy_L40}]{\includegraphics[width=0.5\textwidth]{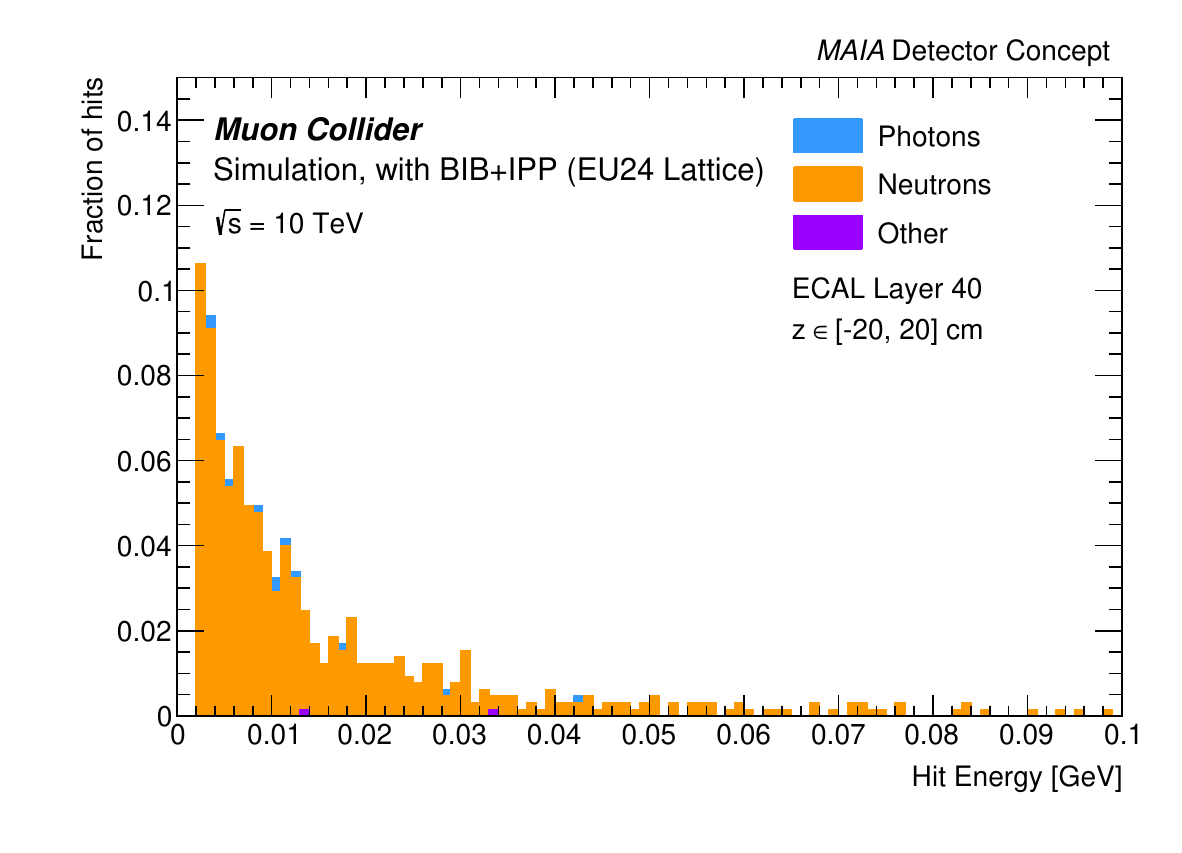}}
 \caption{Composition of the average BIB hit energy spectrum in selected ECAL regions, for different layers of the calorimeter: \protect\subref{fig:ECAL_BIB_energy_L0} 0, 
 \protect\subref{fig:ECAL_BIB_energy_L15} 15,
 and \protect\subref{fig:ECAL_BIB_energy_L40} 40. The colored components highlight the type of particle that originated the shower in the calorimeter cell. The shallower layers are dominated by photon energy deposits, while the deeper layers are dominated by neutrons. The contribution marked as ``Other'' includes muons and other hadrons.}
 \label{fig:ECAL_BIB_energy}
\end{figure}

The BIB particles that reach the calorimeters are mostly photons and neutrons. Figure~\ref{fig:ECAL_BIB_energy} illustrates the BIB energy spectrum and composition in ECAL cells in the central barrel region across different depth layers. In this study, BIB hits in the timing window from -0.5 to 15 ns (relative to expected time of arrival for particles coming from the interaction point) are included in the digitization process, and no further assumptions are made about timing.

In order to mitigate the effects of BIB particles, variable cell thresholds are introduced in the ECAL. The thresholds vary as a function of the polar angle $\theta$ and the calorimeter layer. In the central part of the calorimeter, the diffused BIB photon component becomes negligible and the thresholds can be lowered to target the leftover energy deposition from neutrons.
The application of these variable thresholds further reduces the BIB contribution by an average of 65\%.

\subsection{Muon system}

The muon system is expected to be the least affected by BIB since the upstream calorimeters provide substantial shielding. A notable exception is the very forward region~\cite{Accettura:2023ked}, where the shielding nozzles are located and the BIB backgrounds remain significant.

In the current MAIA design, there is no magnetic field in the muon system, so muons are identified by reconstructing straight tracks matched to extrapolated Inner Tracker trajectories. The layout and baseline technology of the muon system are otherwise kept the same as used in the 3~TeV detector design, although with adjusted dimensions to fit the volume of the other subsystems. The expected detector occupancy and track-finding performance were found compatible with the 3~TeV detector, described in Ref.~\cite{Accettura:2023ked}.

\section{Detector simulation}
\label{sec:simulation}

Detector simulation and reconstruction were performed with the \textsc{MuonColliderSoft} software stack~\cite{Bartosik:2021bjh,Accettura:2023ked} within the Key4hep software ecosystem~\cite{FernandezDeclara:2022voh,Key4hep:2023nmr}. The detector geometry was implemented with the DD4hep detector description toolkit~\cite{dd4hep}, which also supplies the necessary interfaces to Geant4~\cite{Geant4} to simulate the detector response via the \texttt{ddsim}~\cite{ddsim} tool. An important limitation of the simulation setup is that passive materials, such as those required for support structures, cryogenics, and cooling, are not present. The beam pipe, nozzles, and high-density solenoid materials are included in the simulation. Outputs are provided in both the LCIO~\cite{Gaede:2003ip} and the EDM4hep format~\cite{Gaede:2021izq}, which is the standard event data model (EDM) used in Key4hep. 

\begin{figure}[h!]
 \centering
 \includegraphics[width=0.99\textwidth]{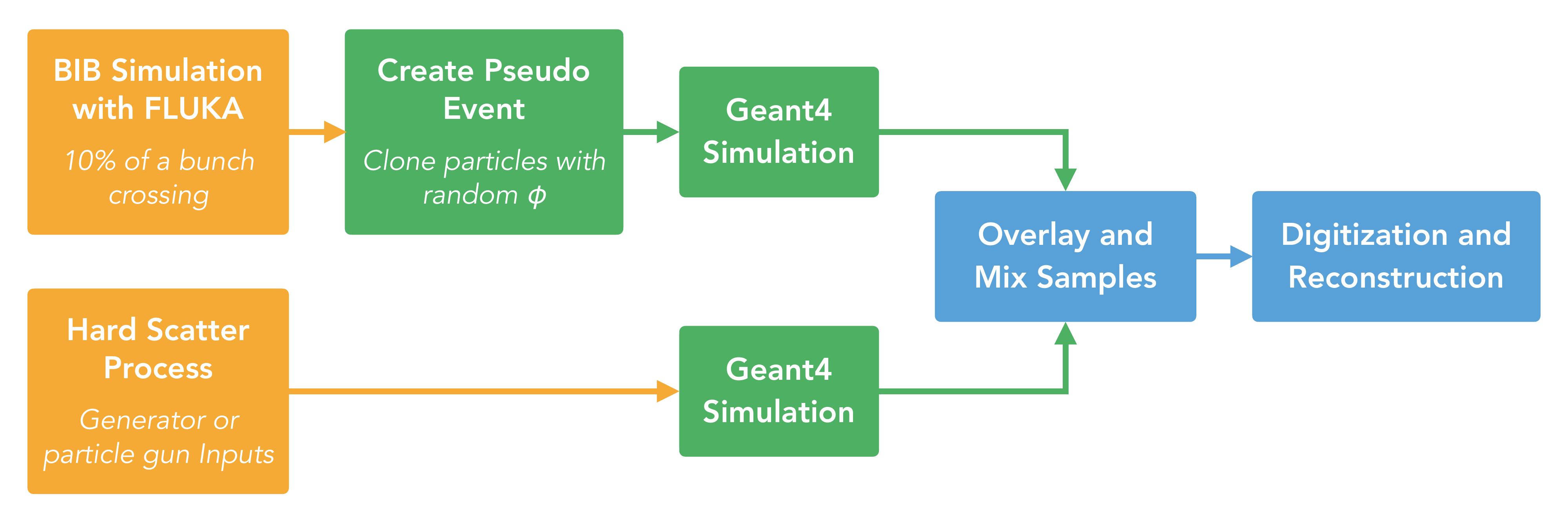}
 \caption{Block diagram of the BIB simulation workflow, summarizing how BIB is generated and combined with hard scattering processes from the \mumu collisions.}
 \label{fig:bib_diagram}
\end{figure}

The BIB simulation is performed with FLUKA, as described in Section~\ref{sec:exp_conditions}. The computational cost of this procedure is considerable; for example, simulating BIB from just a single bunch crossing often takes several weeks to process and requires as much as 50~GB of disk space per event. In order to minimize the usage of computing resources and ensure the statistical independence of each event, a new BIB cloning algorithm was put in place for this study. Here, the total number of BIB particles simulated corresponds to approximately 10\% of a single bunch crossing. 
This ensemble of beam decay particles is then separated into several thousand batches. Within each batch, particles arising from a single beam muon's decay are replicated and rotated with a randomized azimuthal angle, with enough replications that the collection of batches comprises an overlay dataset that corresponds to multiple bunch crossings. A random combination of these batches is then selected to be overlaid with the simulated hits from each physics event.
Each resulting recombined event for overlay is, to a good approximation, statistically independent from the rest. The combined hits from BIB and the physics event are then digitized. The different readout times and resolutions of the different sub-detectors are taken into account by defining integration time windows and selecting hits for the next steps of the simulation. A summary of this process is shown in Figure~\ref{fig:bib_diagram}.

To study the detector performance, samples were generated using a single-particle gun producing either muons, photons, neutrons, or charged pions. Their angles are distributed uniformly in $0 \leq \phi < 2\pi$ and in $8^{\circ} \leq \theta \leq 172^{\circ}$, the latter of which corresponds to $|\eta|<2.44$. The bounds on $\theta$ are selected to avoid hitting the nozzles directly. Photons and neutrons have a uniform energy distribution, with photon (neutron) energies generated from 10~GeV to 5~TeV (10 to 250~GeV). Muons and charged pions have a uniform \pT spectrum, with muon (charged pion) \pT generated from 0.5~GeV to 5~TeV (1~GeV to 1~TeV). The same set of generated events are used for studies both with and without BIB overlay.

In the case of the charged pion samples, the background from incoherent production of $e^+e^-$ is not considered. This choice is not expected to impact the presented results, and a detailed assessment of its impact is left for future work.

\FloatBarrier
\section{Object reconstruction and expected performance}
\label{sec:performance}

Event reconstruction and significant portions of the analysis chain are run using the Gaudi-based event processing framework of Key4hep. 
We also use \textsc{k4MarlinWrapper} to run many existing components from previous studies developed inside the iLCSoft framework~\cite{ilcsoft} using Marlin~\cite{Gaede:2006pj} and LCIO. These functionalities include a combinatorial Kalman filter (CKF)~\cite{Fruhwirth:1987fm,Billoir:1983mz,Billoir:1990vz} based track reconstruction using version 32.1.0 of \textit{A Common Tracking Software} (ACTS)~\cite{Ai:2021ghi} and dedicated overlay and digitization algorithms to handle BIB.

Particle flow event reconstruction is performed using the Pandora algorithm ~\cite{Marshall_2012}, with a configuration adapted from CLIC. The Pandora calorimeter cell clustering approach is based on a calorimeter cone-clustering algorithm, followed by re-clustering algorithms that target specific topologies or optimize a track-cluster match. The large number of calorimeter hits and tracks from BIB increases the computational cost of running Pandora's re-clustering algorithms, so only a select few algorithms are run. The removal of most Pandora re-clustering algorithms significantly reduced the computational cost of event reconstruction without significantly degrading particle efficiencies and resolutions. The following set of Pandora's particle flow object (PFO) identification algorithms are kept: electromagnetic and hadronic showers are differentiated by their cluster topologies, and charged and neutral objects are distinguished through the presence of an associated track.

\subsection{Tracks}
\label{sec:tracking}

As described above, \ref{sec:simulation}, track reconstruction is performed using a CKF
developed for the high occupancy environment of hadron colliders, which is also reasonably well suited to the sizable background present at a muon collider. To mitigate the impact of BIB, hits considered for track reconstruction are required to be consistent with particles produced at the collision. Hit times, corrected by the time-of-flight for a particle traveling at the speed of light, are required to be
within a window of [${-}3\sigma_{t}$, $5\sigma_{t}$] from the beam crossing, where $\sigma_{t}$ refers to the time resolution of the specific tracking subdetector.

Track reconstruction performance is studied with the single muon gun sample, where all muons are produced at the origin. Hits from BIB pose a challenge for pattern recognition and can result in a large rate of fake tracks reconstructed from random combinations of hits. In this study, we present two muon reconstruction working points: Loose tracks are defined as tracks directly from the CKF algorithm, and UltraTight tracks are defined as the subset of Loose tracks that satisfy $p_T > 0.5$ GeV, $n_{hits} \geq 9$, $n_{holes} \leq 3$, $n_{outliers} \leq 4$, and $\chi^2/n_{dof}<3$. This track selection criteria reduce the per-event BIB fake track multiplicity by nearly six orders of magnitude, from $1.6\times10^6$ (Loose tracks) to O(10) (UltraTight tracks). Most of the surviving tracks are in the endcap region. Here, the UltraTight selection is intended primarily as a proof-of-principle working point for BIB fake track rejection, demonstrating that such tracks can be effectively suppressed. A working point optimized for physics performance would likely lie between the Loose and UltraTight selections, achieving a more balanced trade-off between efficiency and fake-track rejection.

\begin{figure}[b!]
    \centering
    \subfloat[\label{fig:eff_vs_theta_10GeV}]{
        \includegraphics[width=0.48\textwidth]{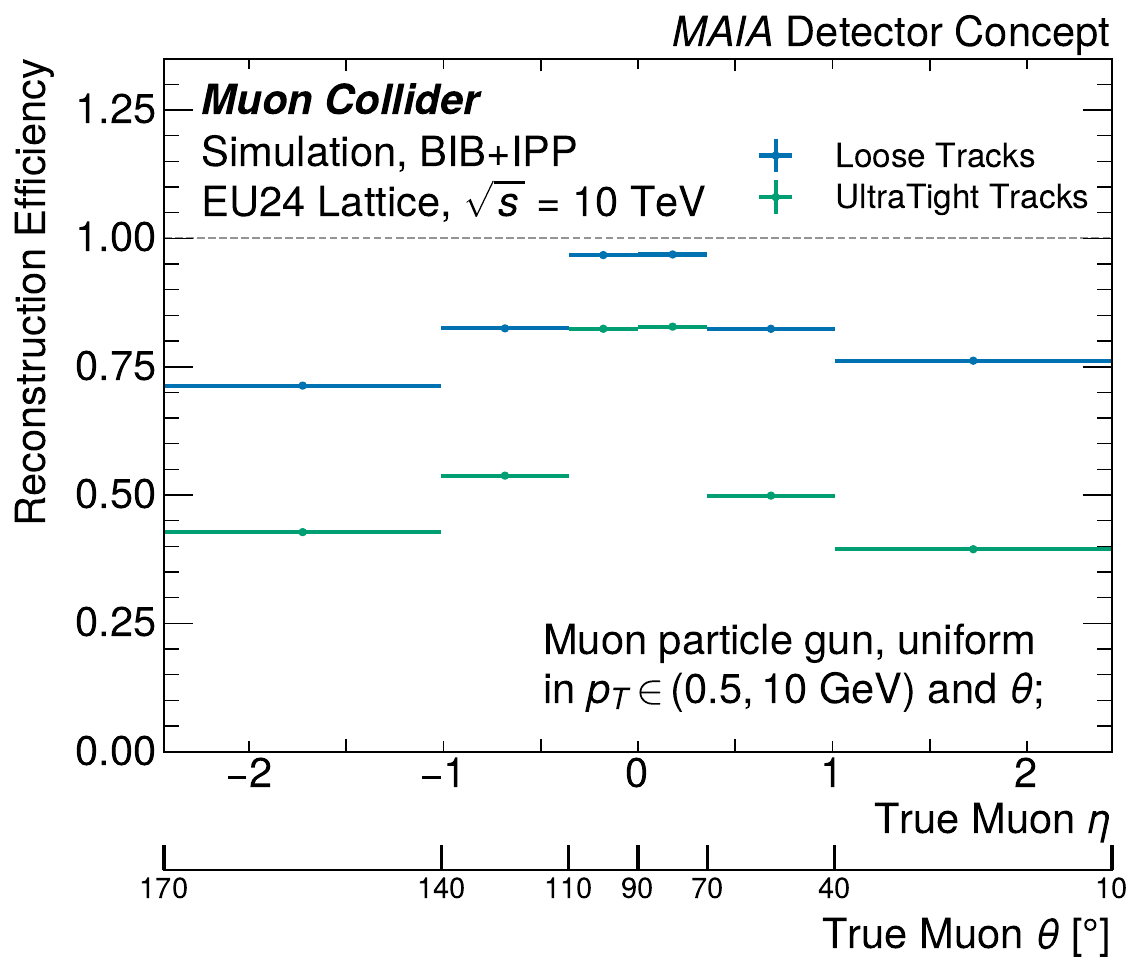}
    }
    \subfloat[\label{fig:eff_vs_theta_5000GeV}]{
        \includegraphics[width=0.48\textwidth]{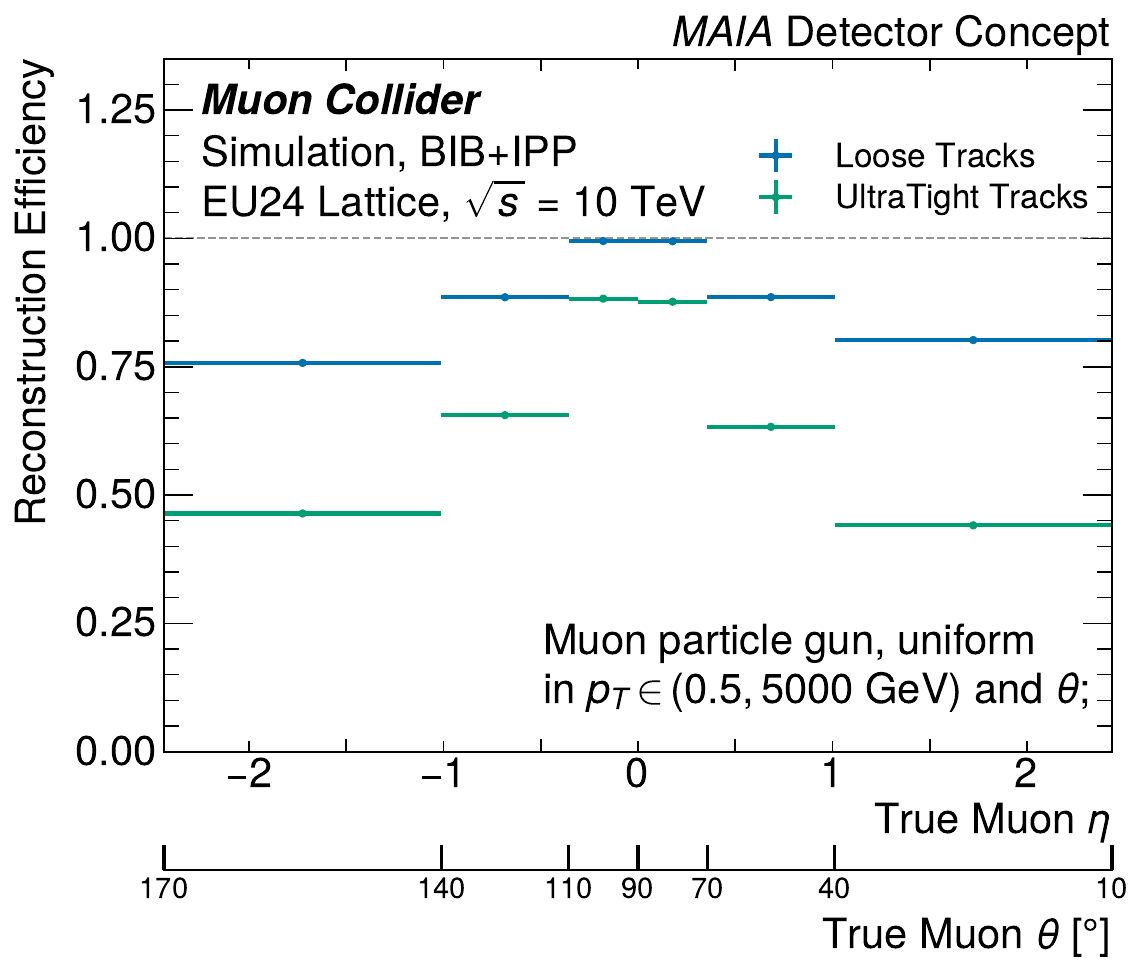}
    }
    \caption{Track reconstruction efficiency as a function of $\eta$ and $\theta$ before
    and after applying the UltraTight selection, comparing tracks with \pT{} between \protect\subref{fig:eff_vs_theta_10GeV}
    [0.5, 10] GeV and \protect\subref{fig:eff_vs_theta_5000GeV} [0.5, 5000] GeV.}
    \label{fig:eff_vs_theta}
\end{figure}

Track reconstruction efficiency with BIB overlaid is presented separately for low and high \pT ranges, and for Loose and UltraTight tracks. Efficiency is defined by matching reconstructed tracks to true muons; for a track to be considered truth-matched, at least 50\% of its hits are required to originate from a muon.
The reconstruction efficiency as a function of $\eta$ and $\theta$ is shown in Figure~\ref{fig:eff_vs_theta}. For tracks traversing only the barrel layers of all three tracking subsystems, corresponding to $57.5^{\circ} < \theta < 122.5^{\circ}$, the reconstruction efficiency evaluated over $0.5 < p_T < 5000~\mathrm{GeV}$ for Loose (UltraTight) tracks reaches 99\% (87\%). In the endcap, the reconstruction efficiency for Loose (UltraTight) tracks drops to $79\%$ ($48\%$). The observed disparities in tracking performance between the central and forward regions highlight the need for further optimization of the tracker endcap geometry, which is left for future work.

\begin{figure}[b!]
    \centering
    \subfloat[\label{fig:eff_vs_pt_barrel_10GeV}]{
        \includegraphics[width=0.48\textwidth]{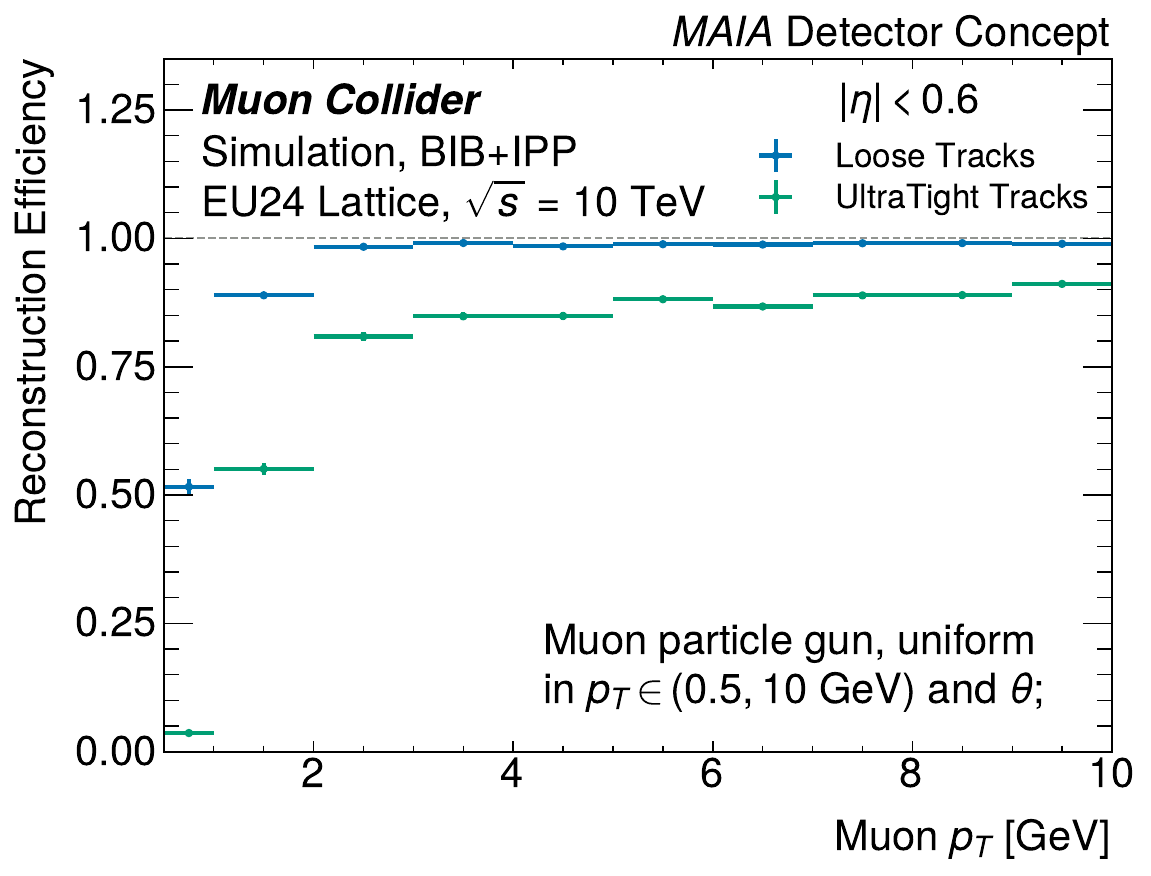}
    }
    \subfloat[\label{fig:eff_vs_pt_barrel_5000GeV}]{
        \includegraphics[width=0.48\textwidth]{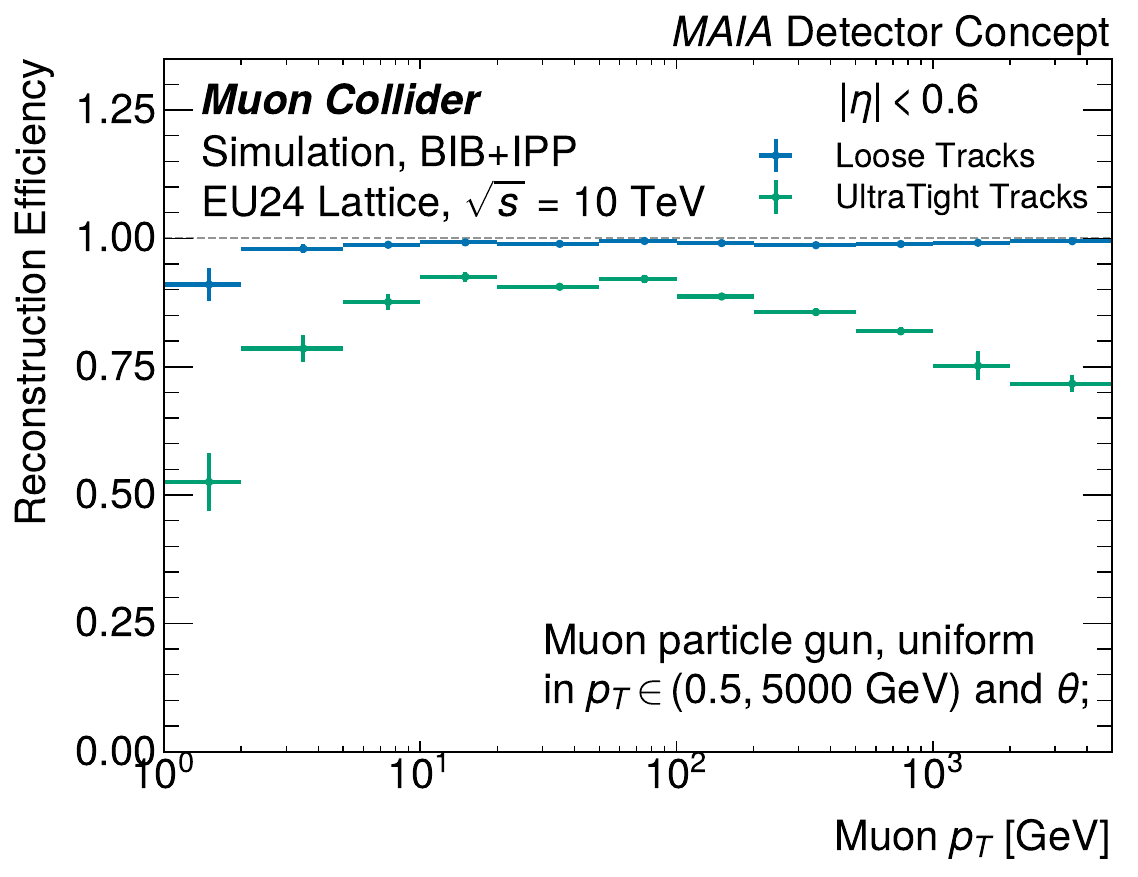}
    }
    \caption{Track reconstruction efficiency in the barrel region as a function of
    \pT{} before and after applying the UltraTight selection, comparing tracks with \pT between \protect\subref{fig:eff_vs_pt_barrel_10GeV}
    [0.5, 10] GeV and \protect\subref{fig:eff_vs_pt_barrel_5000GeV} [0.5, 5000] GeV.}
    \label{fig:eff_vs_pt_barrel}
\end{figure}

\begin{figure}[b!]
    \centering
    \subfloat[\label{fig:eff_vs_pt_endcap_10GeV}]{
        \includegraphics[width=0.48\textwidth]{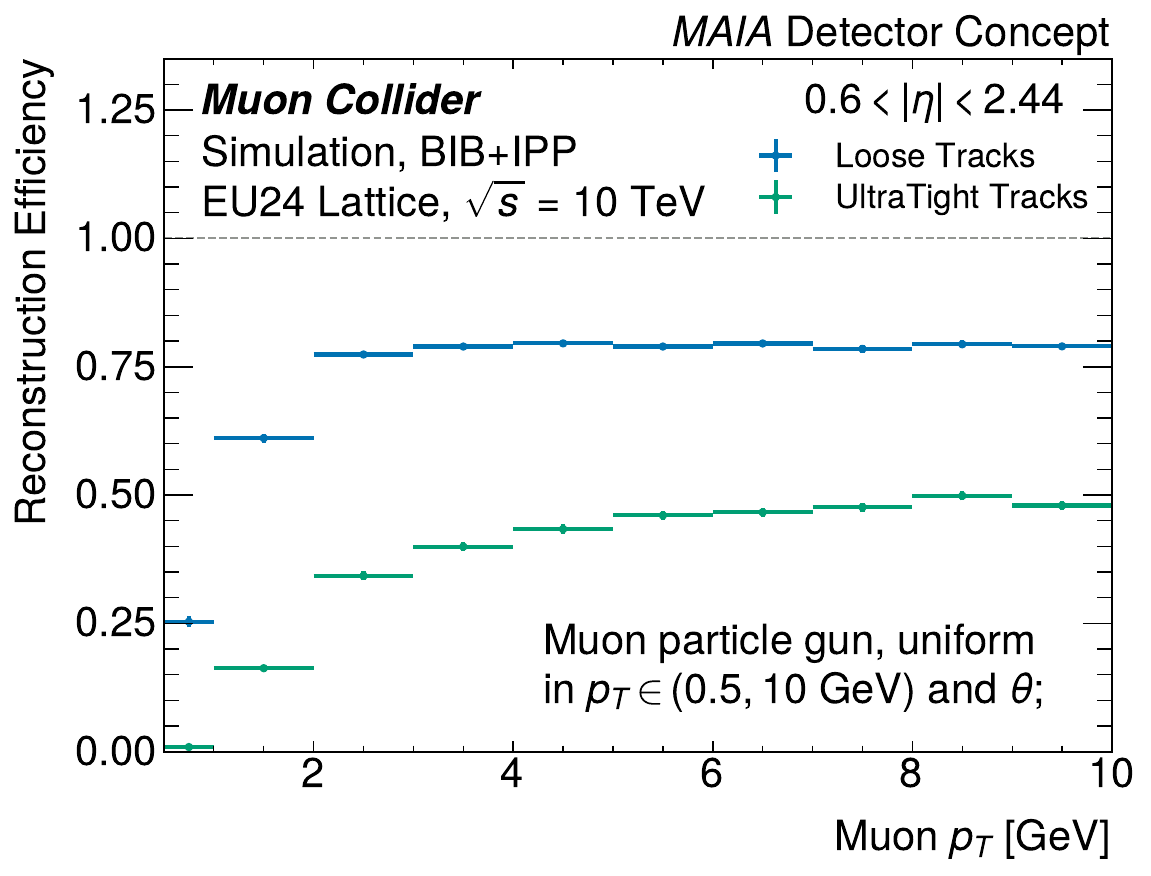}
    }
    \subfloat[\label{fig:eff_vs_pt_endcap_5000GeV}]{
        \includegraphics[width=0.48\textwidth]{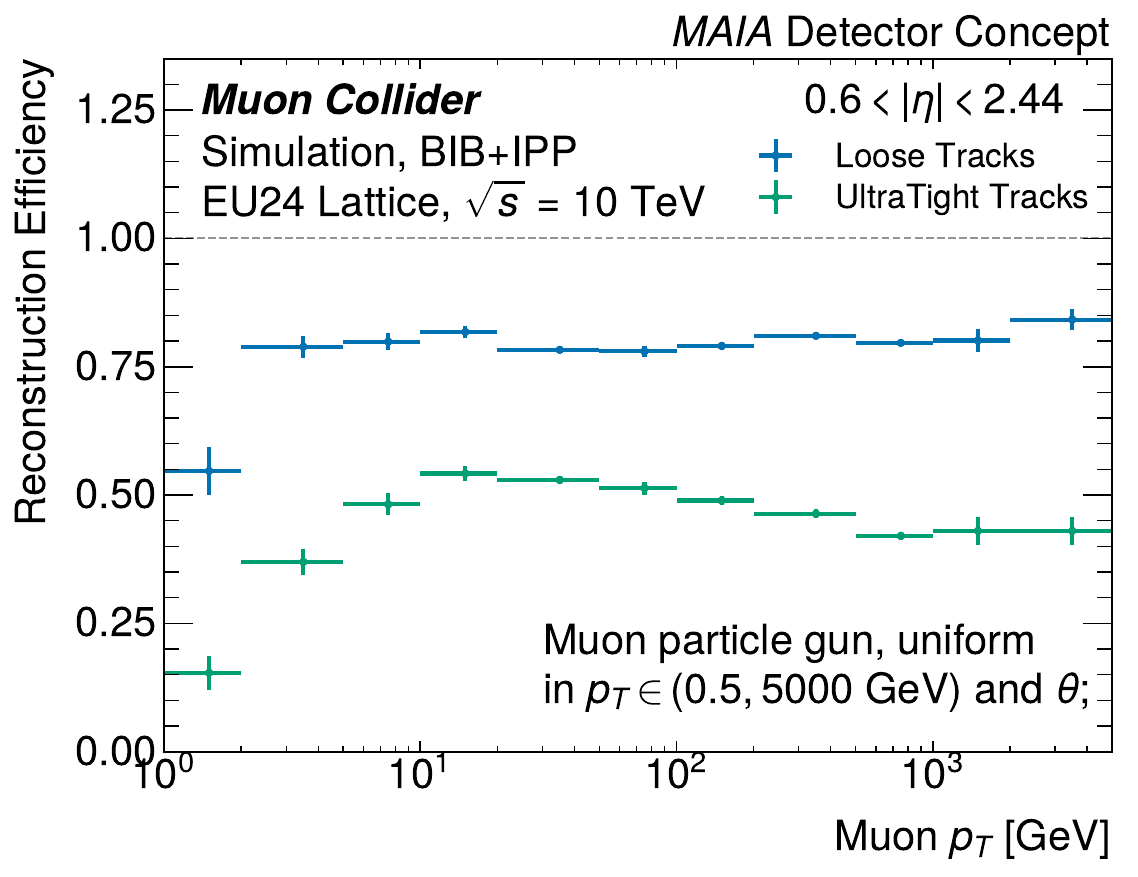}
    }
    \caption{Track reconstruction efficiency in the endcap region as a function of
    \pT{} before and after applying the UltraTight selection, comparing tracks with \pT between \protect\subref{fig:eff_vs_pt_endcap_10GeV}
    [0.5, 10] GeV and \protect\subref{fig:eff_vs_pt_endcap_5000GeV} [0.5, 5000] GeV.}
    \label{fig:eff_vs_pt_endcap}
\end{figure}

Figure~\ref{fig:eff_vs_pt_barrel} and Figure~\ref{fig:eff_vs_pt_endcap} show track reconstruction efficiency as a function of track transverse momentum for the barrel and endcap regions, respectively. Reconstruction efficiency plateaus around 1 GeV, is highest for transverse momenta between roughly 10 and 100 GeV, and drops off at very low and very high transverse momenta.

\begin{figure}[!ht]
    \centering
    \subfloat[\label{fig:pt_res_vs_pt}]{
        \includegraphics[width=0.475\textwidth, valign=t]{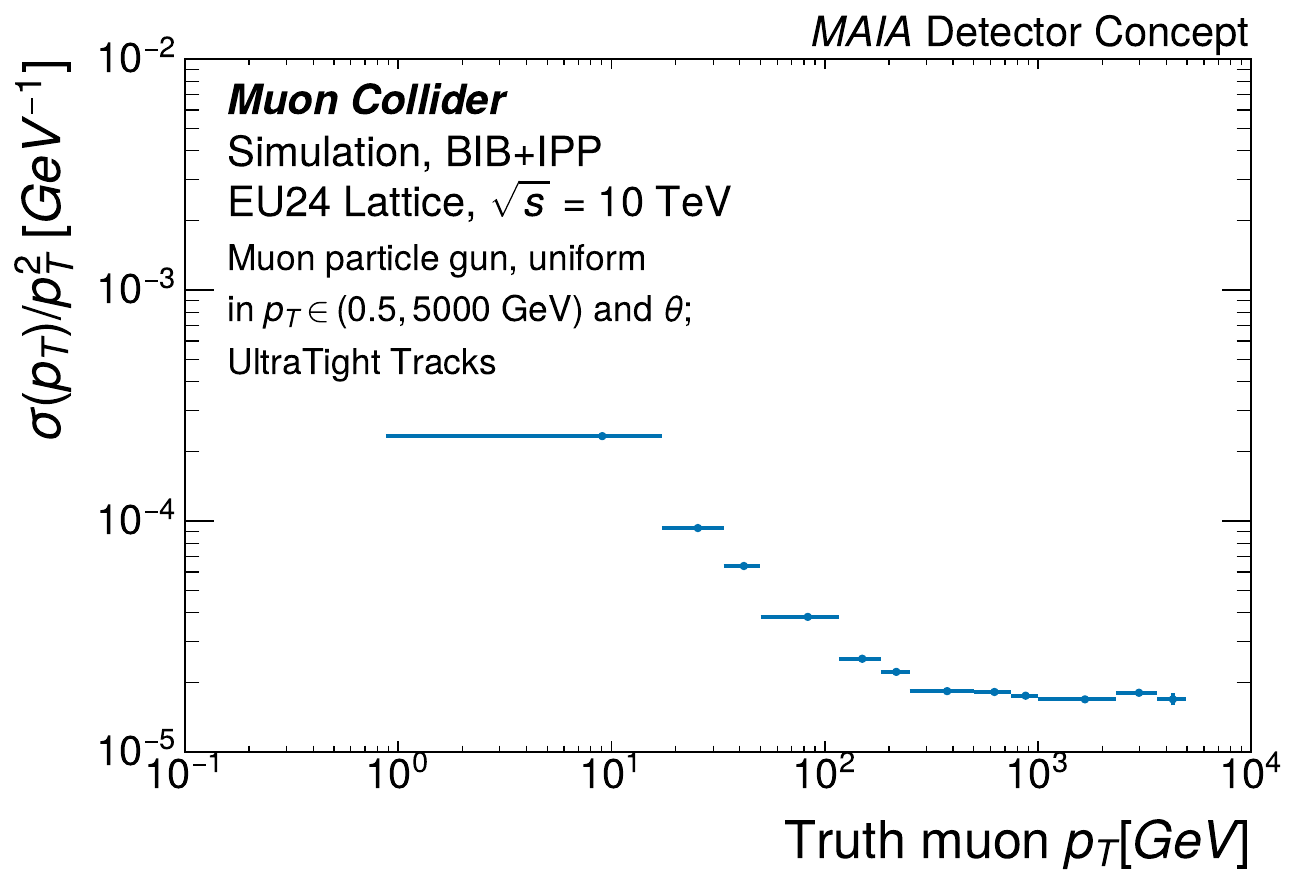}
        \vphantom{\includegraphics[width=0.48\textwidth, valign=t]{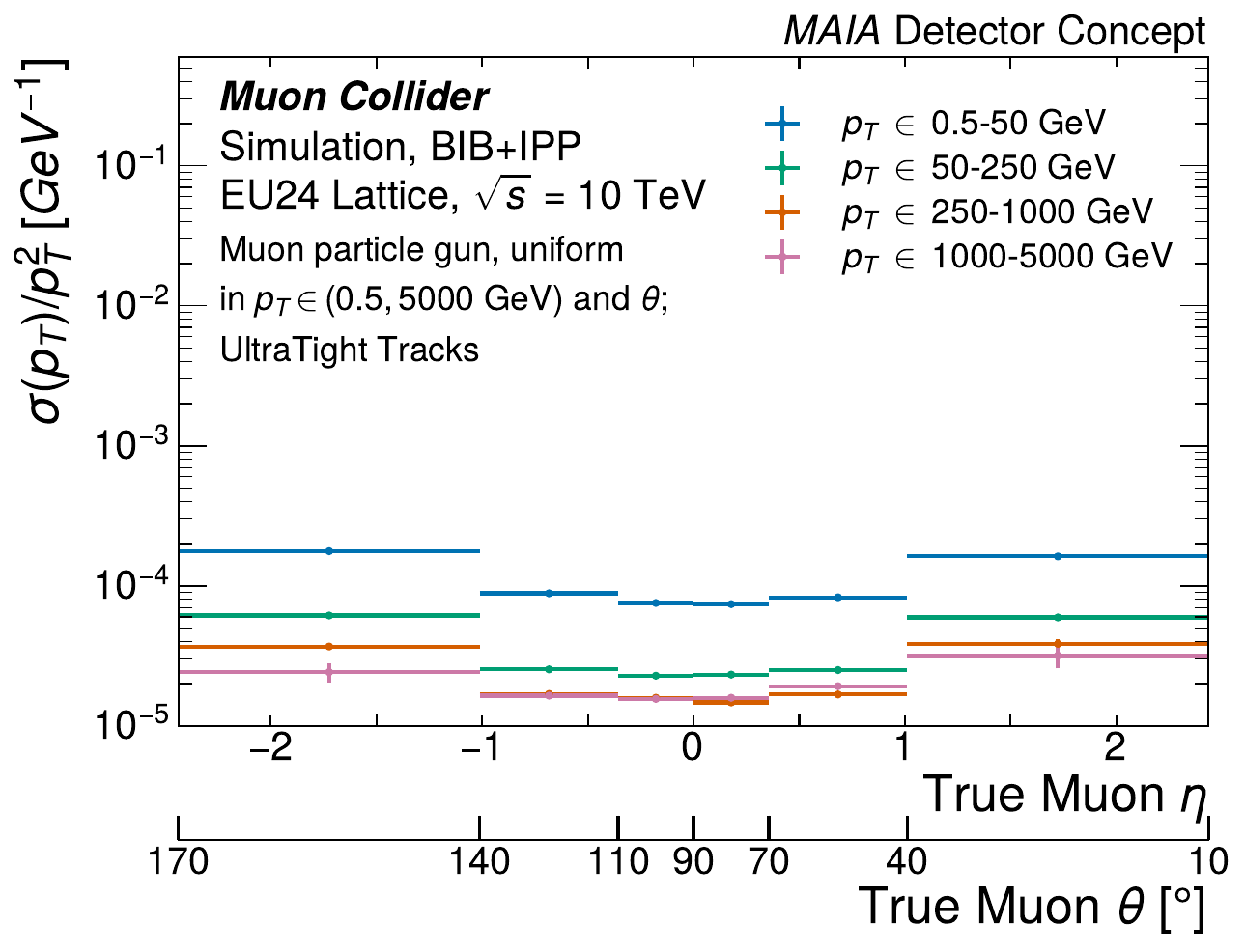}}
    }\subfloat[\label{fig:bib_pt_res_vs_theta}]{
        \includegraphics[width=0.48\textwidth, valign=t]{figures/Tracker/MarkPlots/5000GeVRangePlots/res_pt2_vs_theta_bib.pdf}
    }
    \caption{Inverse track \pT resolution as a function of \protect\subref{fig:pt_res_vs_pt} \pT and \protect\subref{fig:bib_pt_res_vs_theta}
    $\eta$ and $\theta$, comparing four \pT ranges of 0.5-50 GeV, 50-250 GeV, 250-1000 GeV, and 1-5 TeV.}
    \label{fig:pt_res_vs_theta}
\end{figure}

\begin{figure}[t]
    \centering
    \subfloat[\label{fig:d0_res_vs_theta}]{
        \includegraphics[width=0.48\textwidth]{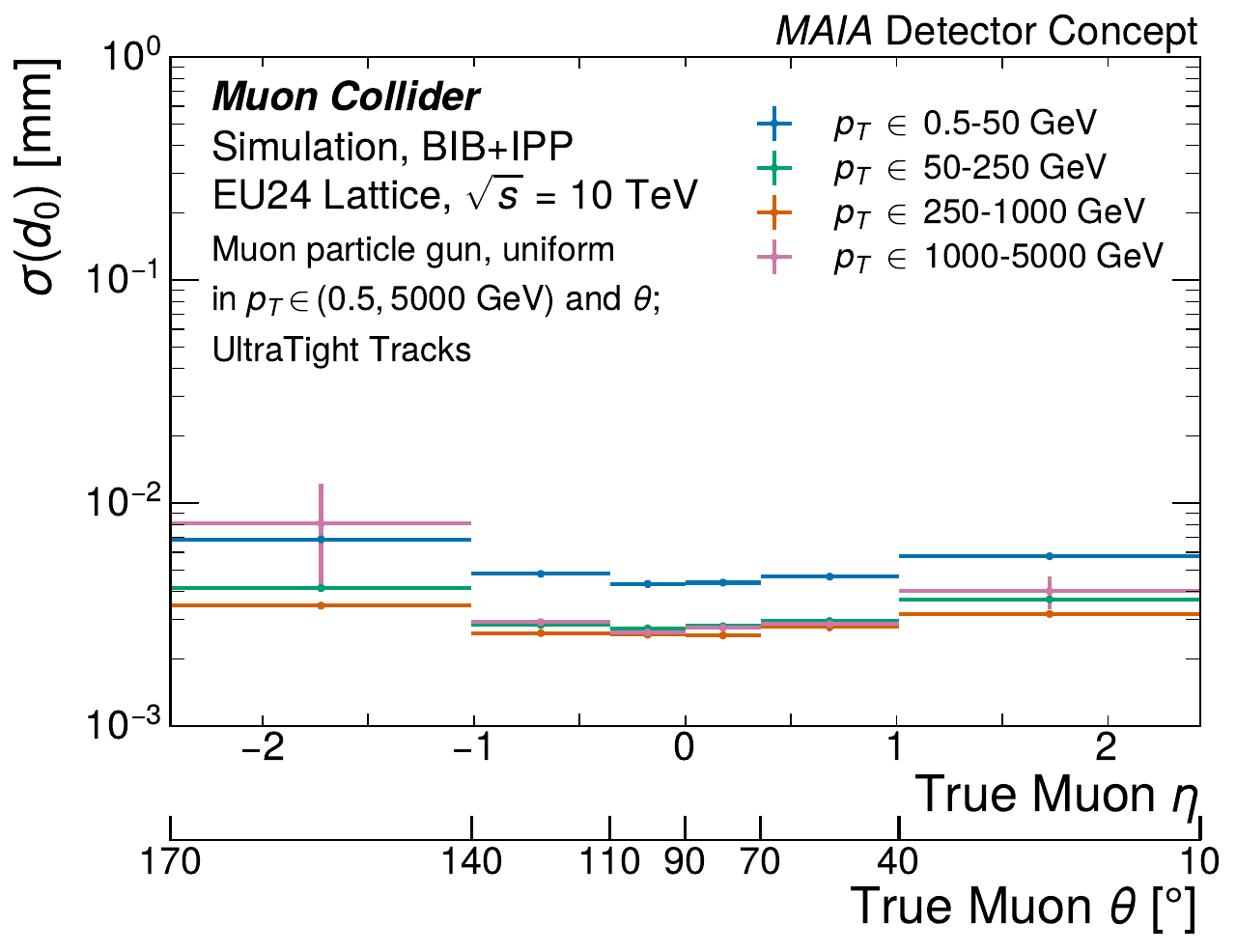}
    }
    \subfloat[\label{fig:bib_z0_res_vs_theta}]{
        \includegraphics[width=0.48\textwidth]{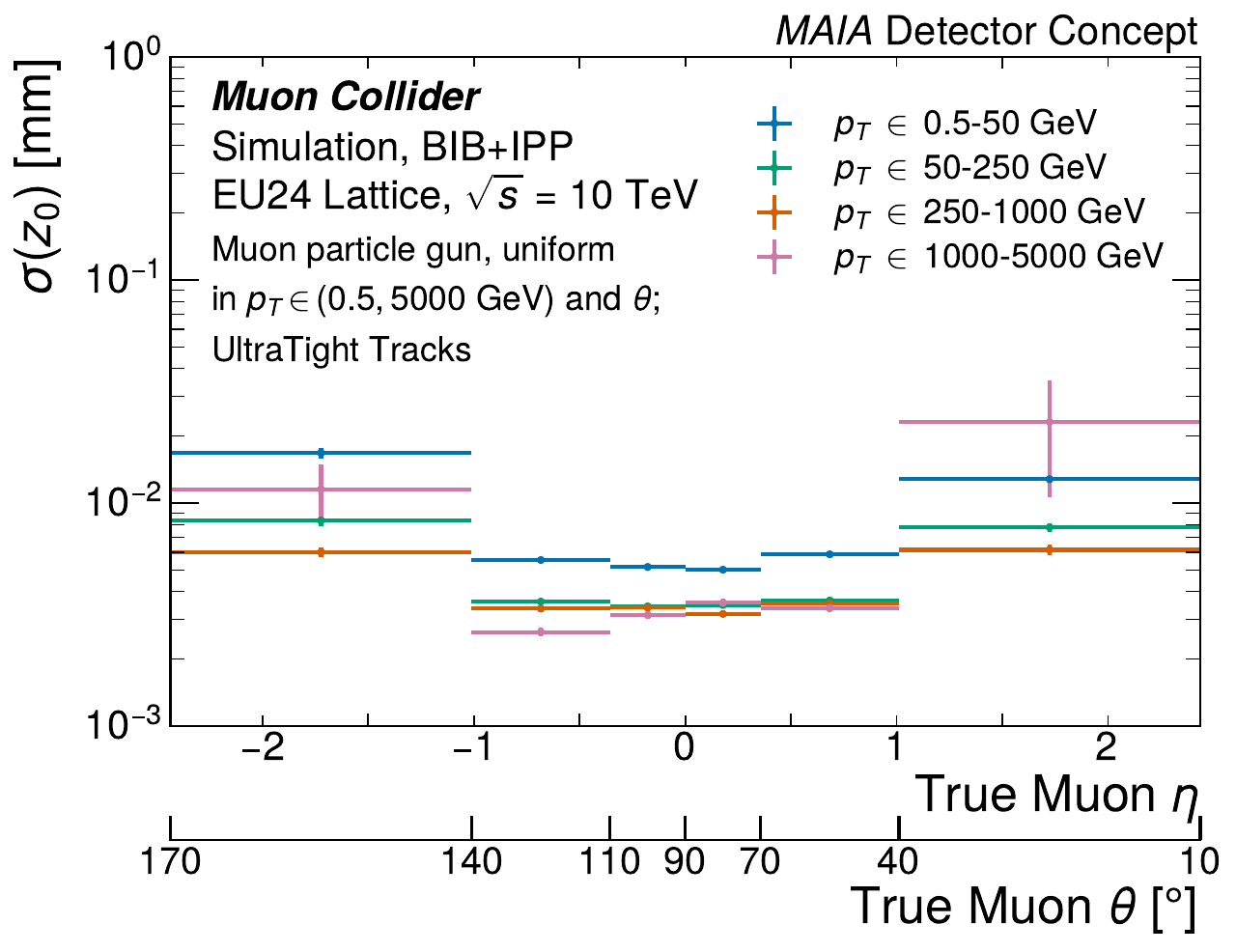}
    }
    \caption{Track \protect\subref{fig:d0_res_vs_theta} $d_0$ and \protect\subref{fig:bib_z0_res_vs_theta} $z_0$ resolution as a function of $\eta$ and $\theta$ comparing four \pT ranges of 0.5-50 GeV, 50-250 GeV, 250-1000 GeV, and 1-5 TeV.}
    \label{fig:impact_res_vs_theta}
\end{figure}

Figure~\ref{fig:pt_res_vs_theta} and Figure~\ref{fig:impact_res_vs_theta} show the transverse momentum and transverse and longitudinal impact parameter resolutions for tracks which pass UltraTight selections, respectively. The inverse transverse momentum resolution is computed by applying a Gaussian fit to residuals between reconstructed and generated \pT. Inverse transverse momentum resolution ranges from $2\times 10^{-4}$ $\text{GeV}^{-1}$ for the lowest \pT sample to $2\times 10^{-5}$ $\text{GeV}^{-1}$ for the highest \pT sample. The impact parameter resolutions are computed by applying a Gaussian fit to the residuals between reconstructed and generated $d_0$ and $z_0$. The resulting impact parameter resolution is found to be around $5~\mu$m in the barrel, and degrades slightly in the endcap.

\subsection{Photons}
\label{sec:photons}

\begin{figure}[!t]
    \centering
    \includegraphics[width=\textwidth]{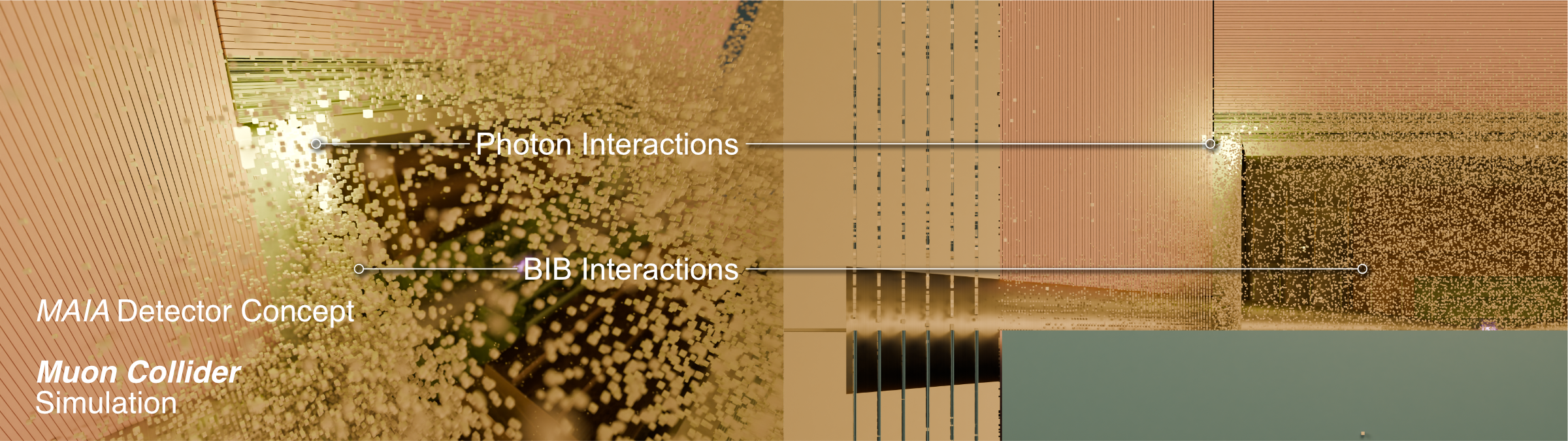}
    \caption{As simulated in GEANT4, detector hits from the shower of a  photon with an energy of 62 GeV are shown in yellow along with hits from BIB. A 3D perspective view (left) and an orthographic $R-z$ projection (right) are shown.}
    \label{fig:photon_eventdisplay}
\end{figure}

Photon reconstruction relies primarily on the measurements made by the ECAL. 
In this study, prior to photon object reconstruction, all hits in a given ECAL cell with timing from -0.5 to 300 ps are summed into a single output. Then, a minimum (maximum) threshold of 0.05~MeV (2.36~GeV) is applied to each cell. %
This upper threshold was chosen to be sufficiently high such that its impact was not discernible. In the case of samples with BIB overlay, this minimum threshold is superseded by higher, region-specific thresholds, as mentioned in Section~\ref{subsec:calorimetry}. An event display of a single photon gun sample overlaid with BIB is shown in  Figure~\ref{fig:photon_eventdisplay}.

As discussed, while BIB particles are predominantly very low energy, their extremely large multiplicity leads to a significant diffuse energy contribution in the calorimeter, particularly in the shallower layers of the ECAL. Unlike the case for track reconstruction, BIB depositions rarely create calorimeter clusters energetic enough to produce a fake a reconstructed particle: using the default Pandora photon reconstruction originally developed for $e^{+}e^{-}$ collider environments, and without applying any photon identification requirements, typical events contain fewer than five fake photon candidates with energies above 20~GeV. However, this BIB contribution can still negatively affect signal photons by adding energy to the clusters and shifting their centroids. Dedicated algorithms that incorporate more sophisticated cleaning techniques and precision timing information could substantially improve both the reconstruction resolution and fake-photon rate. The development and optimization of such algorithms are left for future work.

Photon reconstruction performance was evaluated using a single photon gun sample, both with and without BIB overlay. Here, reconstructed photon matching efficiency was determined by matching the highest-$p_{T}$ Pandora photon candidate within $\Delta R < 0.1$ to the true photon. Only photons with generated energies above 10~GeV are considered. This requirement suppresses a large fraction of BIB particles and reduces ambiguities in the matching procedure, as lower energy thresholds were found to result in more frequent fakes and mismatches.

\begin{figure}[t!]
    \centering
    \subfloat[\label{fig:E_eff_photon}]{
        \includegraphics[width=0.48\textwidth, valign=t, clip]{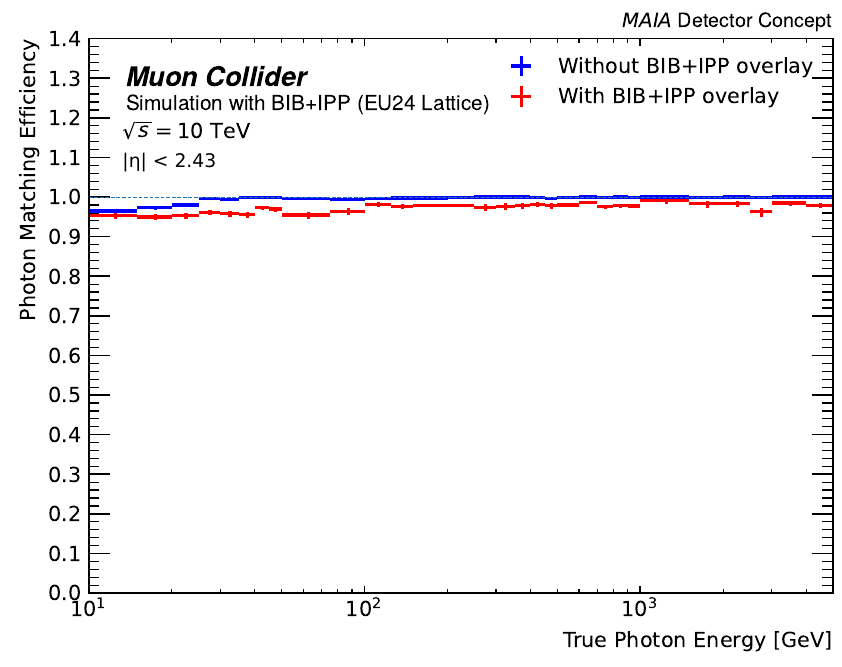}
        \vphantom{\includegraphics[width=0.48\textwidth, valign=t]{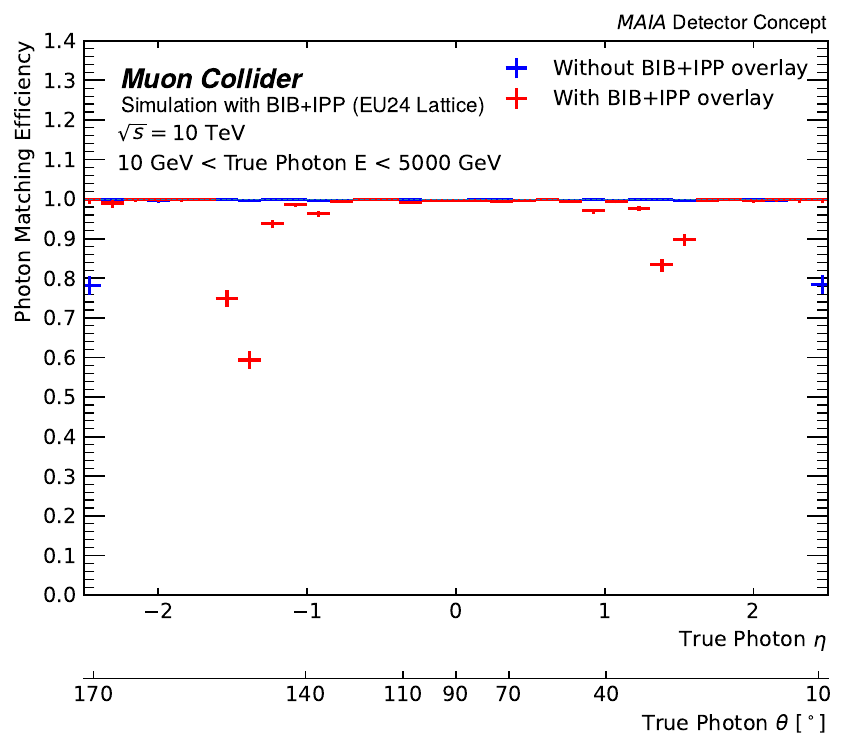}}
    }
    \subfloat[\label{fig:theta_eff_photon}]{
        \includegraphics[width=0.48\textwidth, valign=t]{figures/Photons/v0.8_v8/PhotonEff_BIB_noBIB_v8_maia_eta_deg.pdf}
    }
    \caption{Reconstructed photon matching efficiency with BIB overlay as a function of
    \protect\subref{fig:E_eff_photon} true photon energy and
    \protect\subref{fig:theta_eff_photon} true photon $\eta$ and $\theta$.
    True photons are matched to the photon candidate with highest $p_T$ within $\Delta R \leq 0.1$.}
    \label{fig:photon_eff}
\end{figure}

Photon reconstruction efficiency is shown in Figure~\ref{fig:photon_eff}. Even in the presence of BIB, matching efficiency is high, hovering between 90\% and 95\% across the generated energy range from 10~GeV to 5~TeV. While the clean (no-BIB) efficiency curve is globally higher than the BIB curve, the diverging behavior of the two curves does suggest the presence of some ``fakes", or truth particles spuriously matched to BIB, in the lowest energy bins where the two efficiencies are closest. The matching efficiency as a function of generated photon $\eta$, shown in Figure~\ref{fig:theta_eff_photon}, reveals significant degradation in matching performance in the endcap. The Pandora particle flow reconstruction algorithms and configuration used in this study currently struggle with the significant BIB flux in the forward regions. The notable asymmetry of the degradation in the endcaps is attributable to an underlying asymmetry in the simulated BIB sample and is not a result of the reconstruction process.

\begin{figure}[!htb]
    \centering
    \subfloat[\label{fig:photon_Ereso_split_barrel}]{
        \includegraphics[width=0.48\textwidth]{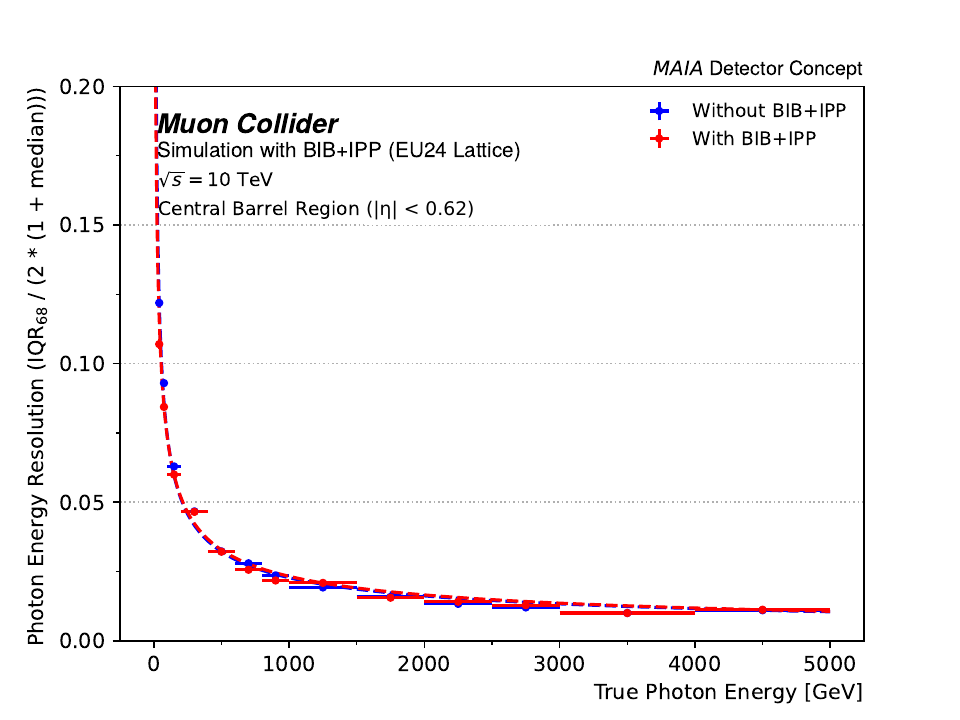}
    }
    \subfloat[\label{fig:photon_Ereso_split_transition}]{
        \includegraphics[width=0.48\textwidth]{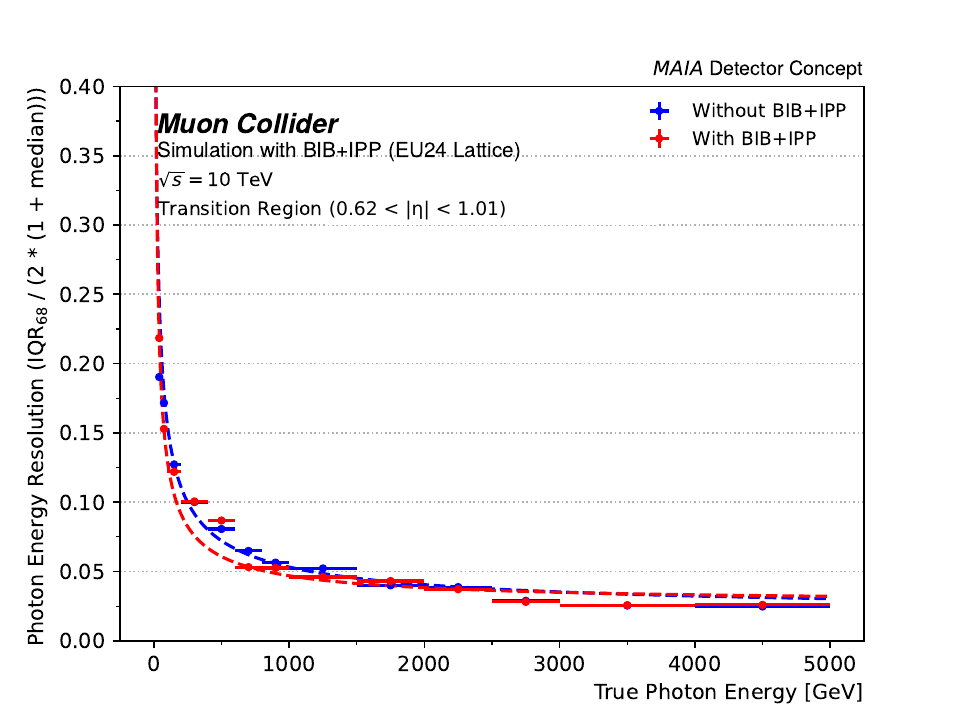}
    }
    \caption{Photon energy resolution of reconstructed photons as a function of true photon energy, split
    into \protect\subref{fig:photon_Ereso_split_barrel} the central barrel and \protect\subref{fig:photon_Ereso_split_transition} transition region.
    Photons in the barrel are best-resolved due to BIB shielding from the solenoid. Data and fitted curves are shown for samples with and without BIB overlay.}
    \label{fig:photon_Ereso_split}
\end{figure}

Energy resolution as a function of generated particle energy, split between two different detector regions, is displayed in Fig ~\ref{fig:photon_Ereso_split}. The resolution value for each energy bin is determined using an interquartile method (IQR$_{68}$) to assess the spread in energy ($(E_{reconstructed}-E_{true})/E_{true}$) in each energy bin. We find the range between the 84th and 16th percentile and scale it by twice the median of the distribution. For a perfectly Gaussian distribution, this value converges to the standard deviation, while remaining robust to outliers. Due to particle mismatching and imperfect energy reconstruction, the energy spread distributions often include negative tails that would significantly bias a Gaussian fit. Error on these data points is calculated using a standard bootstrap method. Resolution curves are fit with a standard function:
\begin{equation}
    \textrm{Res}(E) = \sqrt{\frac{a^2}{E}+\frac{b^2}{E^2}+c^2}
\end{equation}
\label{eq:reso}
where the parameters $a$, $b$, and $c$ correspond to stochastic, noise, and constant terms, respectively.

We split the resolution between two different detector regions in order to demonstrate the effect of the solenoid and the detector geometry on particle reconstruction. The ``central barrel" region ($|\eta|<0.62$) is defined as the region of the ECAL fully shielded by the solenoid, while the ``transition" region ($<0.62<|\eta|<1.01$) covers the ``corners" of the solenoid, including regions with both maximal and minimal solenoid shielding. Particles in the transition region may interact in the barrel detectors, the endcap detectors, or both. Two particles in this region may take very different paths through the solenoid, resulting in different degrees of energy loss from stochastic showering in the magnet, but also different levels of BIB contamination when passing through unshielded regions. The fluctuation in performance introduced by these competing effects is borne out in Figure~\ref{fig:photon_Ereso_split_transition}. In both the barrel and transition regions, the energy resolution curves are similar with and without BIB overlay. In the barrel region, resolution falls below 0.02 (2\%) for multi-TeV photons and is globally below 15\%; resolution is degraded by roughly a factor of two in the barrel region, but still asymptotically reaches 2.5\% at the highest energies. The similarity of the two curves in both regions indicates that the addition of the BIB does not have a significant negative impact on ECAL performance (other than in the forward regions).
The performance in the forward region was not yet quantified. Dedicated algorithms targeting this exceptionally challenging region are in development and will be documented in future work.

\subsection{Neutral hadrons}
\label{sec:hadrons}

Neutral hadron reconstruction relies primarily on the HCAL, with some contributions by the ECAL depending on particle energy and position. Here, the same cell timing and energy threshold requirements described for the ECAL in Section~\ref{sec:photons} are applied to the HCAL. 

\begin{figure}[b!]
    \centering
    \includegraphics[width=\textwidth]{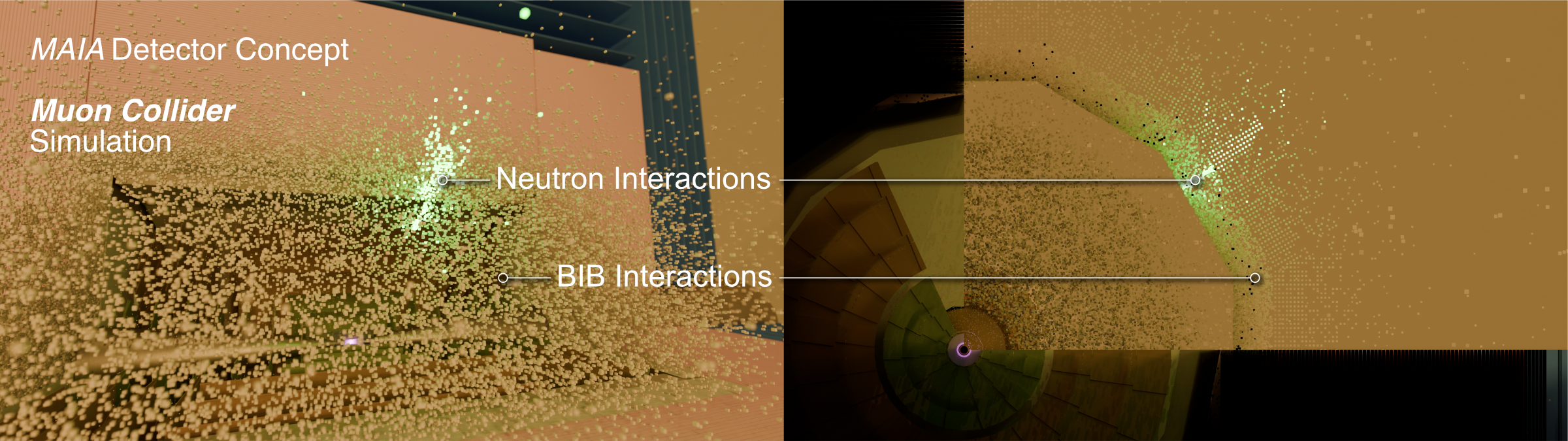}
    \caption{As simulated in GEANT4, detector hits from the shower of a neutron with an energy of 73 GeV are shown in green along with hits from BIB. A 3D perspective view (left) and an orthographic $x-y$ projection (right) are shown.}
    \label{fig:neutron_eventdisplay}
\end{figure}

The characterization of the HCAL performance is accomplished by studying the reconstruction efficiency and energy resolution of single neutron particle gun samples. Figure~\ref{fig:neutron_eventdisplay} shows a 73~GeV neutron shower event display, where neutron cell hits in green and BIB hits in brown.

Neutron clusters are reconstructed and identified using Pandora's particle identification algorithms.
Since this implementation of Pandora clustering is not optimized for a muon collider environment, soft clusters are highly contaminated with BIB energy and cannot be discerned from the BIB background. Therefore, particle flow objects with $E_{PFO} < 10$~GeV are rejected, and all studies require a cut of 20~GeV on the generated particle energy, $E_{true}$. The $E_{true}$ cut limits the low-energy biasing created by a PFO energy cut. %

\begin{figure}[b!]
    \centering
    \subfloat[\label{fig:neutron_eff_energy}]{
        \includegraphics[width=0.48\textwidth, valign=t, clip]{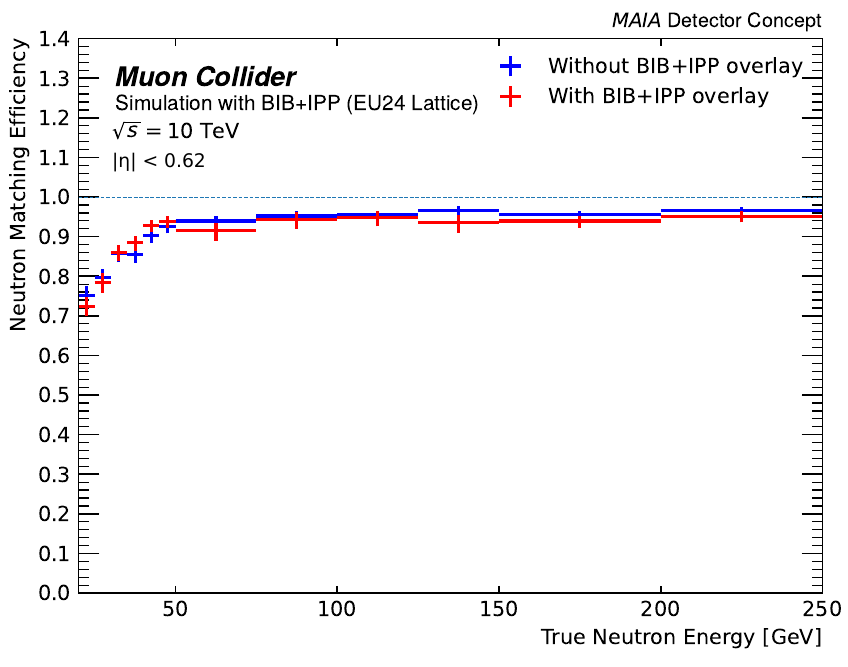}
        \vphantom{\includegraphics[width=0.48\textwidth, valign=t]{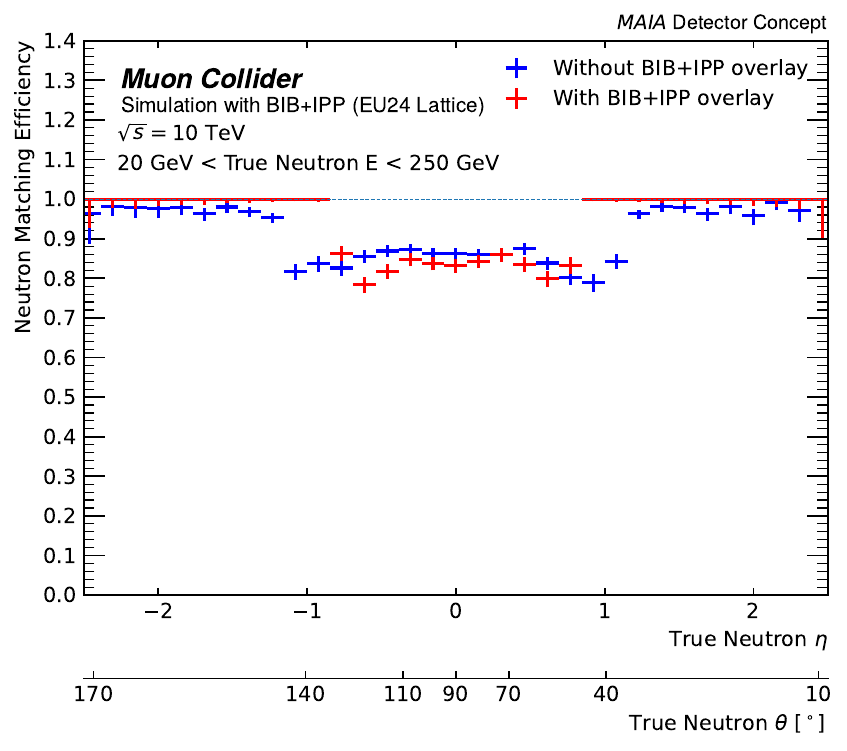}}
    }
    \subfloat[\label{fig:neutron_eff_theta}]{
        \includegraphics[width=0.48\textwidth, valign=t]{figures/Neutrons/v0.8_v8/NeutronEff_BIB_noBIB_v8_maia_eta_deg.pdf}
    }
    \caption{Neutron reconstruction efficiency as a function of \protect\subref{fig:neutron_eff_energy} true MC particle energy and \protect\subref{fig:neutron_eff_theta} true MC particle $\eta$ and $\theta$. Efficiency vs energy is restricted to the barrel region to exclude spurious high-energy matching in the endcaps.}
    \label{fig:neutron_eff}
\end{figure}

Reconstructed neutron matching efficiency is displayed in Figure~\ref{fig:neutron_eff}. The matching process for neutrons closely parallels the photon procedure: we match the truth particle to the neutron-identified PFO with the highest transverse momentum within $\Delta R<0.4$, where the cone requirement has been widened to accommodate the softer structure of neutron showers. In Figure~\ref{fig:neutron_eff_energy}, the perfect efficiency in the endcap regions is indeed due to spurious matching, or ``fakes". The enormous flux of BIB in the HCAL endcaps causes a large portion of the detector to ``light up," creating a single highly-energetic cluster that the current Pandora algorithms are unable to resolve into individual particles. The reconstruction algorithm automatically matches the true particles to these multi-TeV Pandora clusters, leading to a 100\% spurious efficiency in the endcaps. Therefore, we exclude the endcaps in Figure~\ref{fig:neutron_eff_energy}, allowing a more accurate assessment of the matching efficiency energy dependence. Matching efficiency remains high, between 90 and 100\%, for $E_{true}\gtrsim50$~GeV, both with and without BIB overlay. At lower energies, the efficiency is slightly higher for the BIB sample, indicating either lingering spurious matching or that BIB contamination enhances the energy and angular spread of the original reconstructed particle, leading to a match only in the BIB scenario.

\begin{figure}
        \centering
        \includegraphics[width=0.6\textwidth]{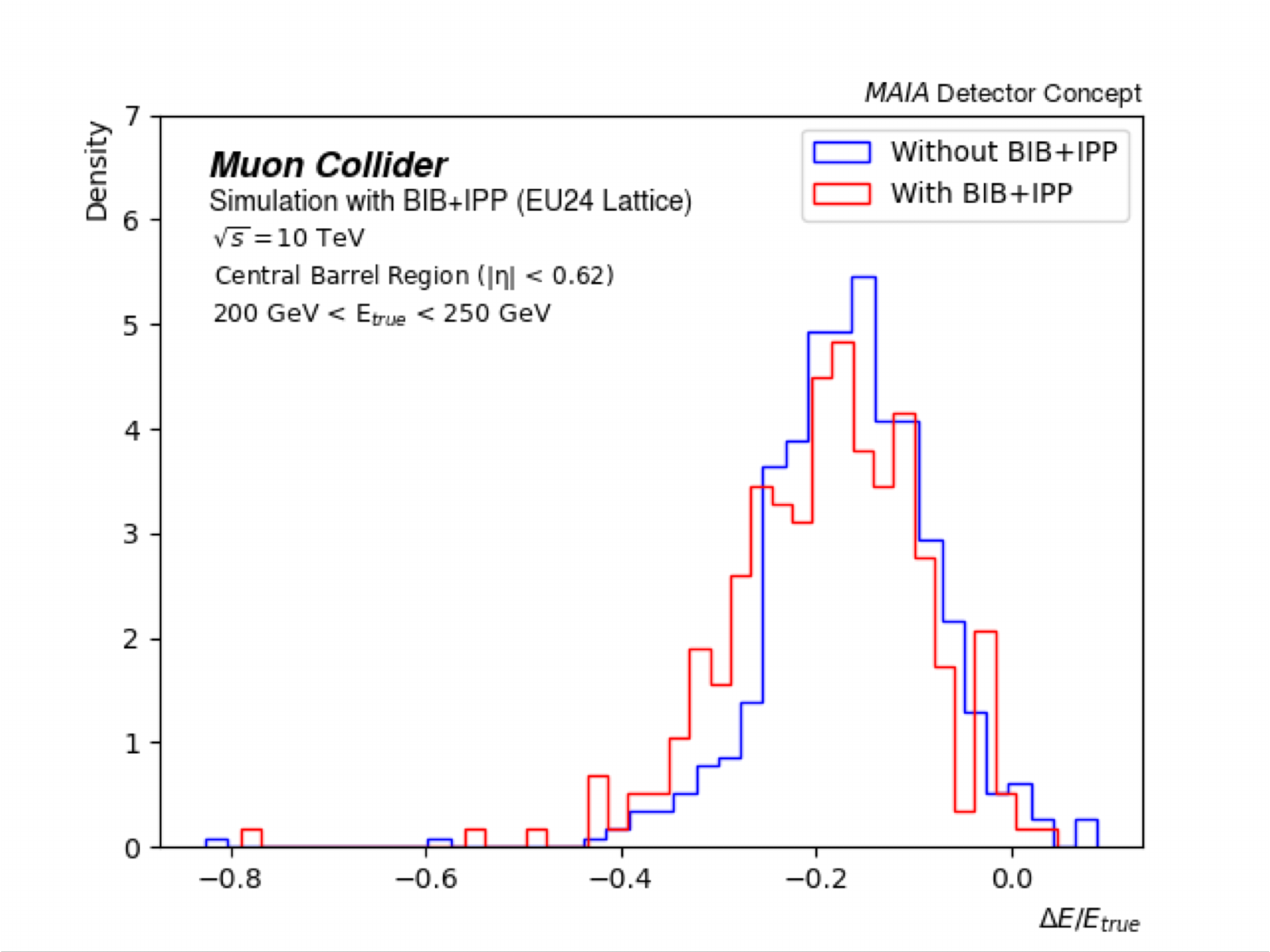}
    \caption{Energy resolution (IQR) for simulated neutrons with 200 GeV $<E_{true}<$ 250 GeV, with and without BIB overlay. IQR$_{68} = 0.090\pm0.001$ without BIB, and IQR$_{68} = 0.108\pm0.002$ with BIB.}
    \label{fig:neutron_gaussians}
\end{figure}

Energy resolution for reconstructed neutrons is assessed with the same IQR$_{68}$ metric that is used for photons. Example resolution distributions of $\Delta E / E_{true}$ both with and without BIB are shown in in Fig. \ref{fig:neutron_gaussians}. We choose these distributions to represent the resolution of the HCAL, since soft (low-energy) neutrons deposit a significant fraction of their energy in the ECAL. Therefore, the energy resolution for low-E neutrons is affected by the respective calibrations of the ECAL and HCAL, which makes it more representative of the interplay between the two subdetectors than of the HCAL performance itself. Moreover, we focus only on the central barrel region, due to the challenges presented by high BIB flux in the endcaps, as discussed. 

While the distributions displayed in Fig. \ref{fig:neutron_gaussians} are qualitatively similar, the addition of BIB overlay degrades resolution slightly. Additionally, both distributions are shifted from 0, indicating that energy is globally underestimated by roughly 20\%. This offset median provides further motivation for an optimized energy calibration of the HCAL.

\subsection{Charged Pions}
\label{sec:pions}

The reconstruction of charged pions requires the combination of tracking and calorimeter information in order to form a high-quality track-to-cluster match. The selected re-clustering algorithms run by Pandora aim to form a cluster that contains energy consistent with a nearby track. Requirements on the tracks, most notably on its momentum uncertainty, are thereby enforced to ensure a proper track momentum to cluster energy comparison. 

The final track-cluster matching is performed by matching a cluster to a high-quality track within 200~mm from the propagated track position at the ECAL face. In the case of multiple nearby clusters, the one that is closest and has the best energy match is selected. To reject high-quality tracks matching to low-energy cluster fragments or BIB clusters, an additional requirement is enforced on the match: $(p_T^{track} - E^{cluster}) / p_{T}^{track} < 0.95$. The highest \pT charged PFO within $\Delta R < 0.1$ is matched to the generated charged pion. The track-cluster matching efficiency is defined as the number of particle gun events containing a matched charged PFO divided by the number containing a reconstructed track, as defined in Section \ref{sec:tracking}.

Pandora identifies all charged PFOs that fail both an electron and muon identification algorithm as charged pions. The electron identification algorithm compares the charged PFO cluster topology to an expected EM shower shape, and the muon identification algorithm discriminates showers that are consistent with that from a minimum ionizing particle. The charged pion identification efficiency is defined as the number of particle gun events containing an matched charged pion divided by the number containing a reconstructed track, as defined in section \ref{sec:tracking}. 

\begin{figure}[!ht]
    \centering
    \subfloat[\label{fig:pion_without_bib}]{
        \includegraphics[width=0.48\textwidth]{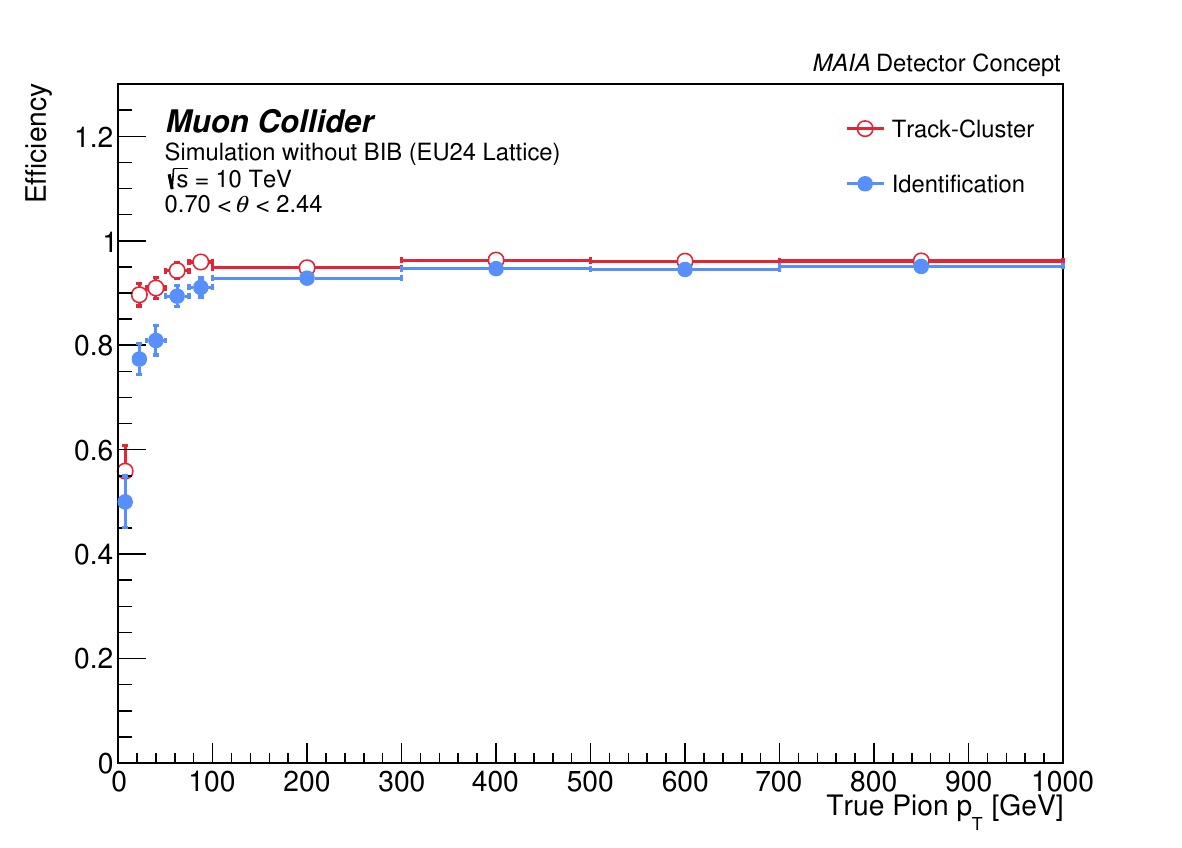}
    }
    \subfloat[\label{fig:pion_with_bib}]{
        \includegraphics[width=0.48\textwidth]{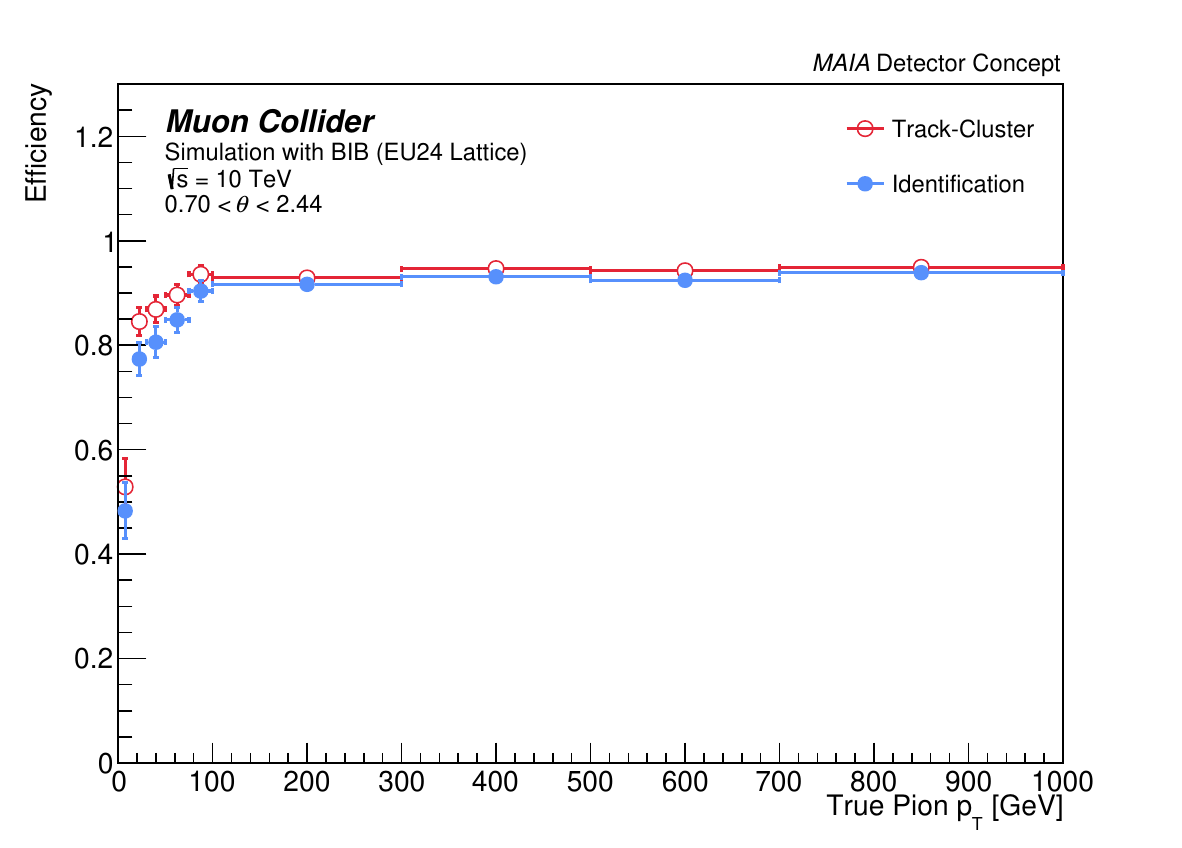}
    }
    \caption{Track-cluster matching and PFO identification efficiencies for charged pions in the calorimeter barrel region, \protect\subref{fig:pion_without_bib} without and \protect\subref{fig:pion_with_bib} with BIB overlay. The efficiencies are defined with respect to the tracking efficiency, as provided in section \ref{sec:tracking}.}
    \label{fig:chargedPion-trk-cls-pfo}
\end{figure}

The track-cluster matching and PFO identification efficiencies are provided for the calorimeter barrel region in Figure \ref{fig:chargedPion-trk-cls-pfo}. The track-cluster matching and PFO identification efficiencies are lowest in the low-\pT regime, a result from tracks failing the requirements to be considered for a track-cluster match, tracks matching to cluster fragments, and tracks matching to clusters originating from BIB. These effects become less prominent at higher \pT, where the track-cluster matching and PFO identification efficiencies approach 95\%.

\FloatBarrier
\section{Future work}
\label{sec:future}

This conceptual detector design can be used as a foundation for further study, particularly for higher-level object reconstruction and physics projections in the presence of BIB and SM collision backgrounds. The performance of this detector is expected to increase with further optimization of the detector geometry, subdetector technology choice, and reconstruction algorithms. As the details of the accelerator lattice, particularly the final focus and length of the straight sections leading to the experimental area, change, the character and level of the BIB is expected to change. Such changes may require additional reoptimization of the detector design as needed.

The optimization of the nozzle geometry, material, and structure remains a major area for improvement. Small changes to the geometry or material makeup of the nozzle can have significant effects on occupancy of the detectors, particularly for the region closest to the collision. Coarse optimizations have been performed. However, running BIB simulation through iterative nozzle designs is computationally intensive, and major improvements in computing may be crucial in efficiently exploring the full space of nozzle designs. The use of machine learning, such as generative networks to more efficiently simulate interactions with the nozzle~\cite{Paganini:2017dwg}, may be valuable in this optimization.

We plan to explore major design changes to the detector as well, especially when it comes to the forward region. The details of the endcap geometry and its interplay with the barrel region have not been be the focus of this work and will be optimized in the future. The placement of the transition region will depend heavily on the character of the BIB, the energy of the collisions, and the physics processes of interest.

The overall layout of the entire silicon tracker can also be reoptimized. We aim to explore the use of double layers in the Inner Tracker and Outer Tracker designs paired with a redesign of the overall geometry, potentially using machine learning to optimize the many parameters of the tracker layout. Double layers have been studied in the past in the context of a muon collider experiment, but we aim to study their utility in the outer layers of the tracking system. In the tracker endcaps, we will explore the use of tilted geometries to reduce the average angle of incidence, making better use of the high spatial resolution of the modules and reducing the average cluster size.

Because the muon system is mostly unaffected by BIB, the design of the muon system and muon reconstruction -- in particular the matching of tracks between the muon system and charged particle tracker -- has not been explored in detail thus far. The current muon system largely operates as
a tagger, without an additional magnetic field to form a spectrometer. We aim to study alternate
designs in the future, as well as the use of HCAL technologies that use micropattern gaseous detectors~\cite{Pellecchia:2024bpr} with single-particle tracking abilities to simultaneously act as an HCAL and a muon tagger. Such a technology could be used in tandem with the iron-scintillator design presented here for a hybrid HCAL system.

A dedicated feasibility study is required for the solenoid. In these studies, we have assumed a 5T magnet with an inner radius of 150~cm simulated as a region of high-density materials, aluminum and steel. The feasibility of such a magnet needs to be evaluated, particularly given the sensitivity of the calorimetry to the amount of material in the magnet. A more realistic model of such a large high-field magnet should incorporate radiation tolerance requirements, and the structural and material design must be able to handle the large mechanical stresses from the field.

A goal of such an experiment is to be able to run without on-detector buffering. The feasibility of this goal heavily depends on the machine parameters, the level of background, and the ability to read out the detector at the full collision rate. Preliminary studies suggest that such a readout scenario may be possible with this concept, but more mature data throughput studies would be required to demonstrate this feasibility. One potential area of concern is the vertex layers of the silicon tracker, which would see the highest occupancy rates overall and require the most bandwidth for readout.

The algorithms used in the results presented here have been coarsely optimized for reasonable performance. The performance is expected to greatly benefit from additional algorithmic optimizations for many of the physics objects discussed here. Improved photon reconstruction in the presence of BIB is crucial for many of the of the physics goals at such a machine, and reconstructed photons that arise purely from BIB contributions are not yet studied here. Since the high-dimensional shower shape in the ECAL is the only handle for photons, photon reconstruction is an ideal case for machine learning discriminators given the large effect of BIB on the ECAL. The optimization of the tracking algorithm may need to be reevaluated should further changes to the detector lattice cause significant changes in the backgrounds expected in the tracker.

We also leave the study of higher level physics objects for future work. Flavor tagging of jets will serve as an important tool, particularly for the program of studying the Higgs boson and new physics that couples to heavy flavor. Flavor tagging of jets requires a careful interplay of tracker and calorimeter measurements and will be better understood as the more fundamental reconstruction algorithms improve in sophistication. The same is true for the identification of hadronically decaying $\tau$ leptons.
Optimizing a general particle flow algorithm for the unique muon collider environment will require
significant high-level reconstruction studies once basic objects are very well understood.

In addition to these lower-level improvements, studies of the overall physics performance are necessary, especially in the presence of a sophisticated suite of backgrounds. Studies of the multi-object mass resolutions, missing energy measurement performance, and primary and secondary vertex reconstruction are crucial to ensuring robust physics performance from the experiment. Dedicated studies of full signal processes in the presence of realistic backgrounds are required to characterize the reach for standard model measurements and in searches for physics beyond the standard model.

\section{Conclusions}
\label{sec:conclusions}

The MAIA detector concept shows considerable promise in advancing our capability to conduct precision and discovery physics at a 10~TeV muon collider. Here, track, photon, neutron, and charged pion measurements showcase the reconstruction capability of the tracker and calorimeter systems: under realistic BIB conditions, we observe minimal degradation in performance for energetic particles in the central region of the detector. 

Track reconstruction performance with BIB is evaluated at two working points, the tightest of which successfully rejects nearly all fake tracks, albeit with a reduction in efficiency that remains to be improved. Nevertheless, tracks passing these stringent selections achieve transverse momentum and impact parameter resolutions approaching 1\% and 2~$\mu$m, respectively. For photons with E~$>$~10~GeV in the presence of BIB, reconstruction efficiencies surpass 95\% throughout the central barrel, with energy resolutions approaching 1\%. Energetic charged pions (E~$>$~100~GeV) across the detector feature reconstruction efficiencies approaching 95\%. A critical driver of this performance is several ps-scale timing resolutions and high-granularity calorimetry.

This work establishes a baseline detector concept for physics studies at a 10~TeV muon collider that can serve as a foundation for future detector optimization and physics performance studies. Key areas of future work include optimization of the nozzle geometry, tracker layout, and forward detector regions, as well as detailed studies of the magnet design and readout architecture. Continued development of reconstruction algorithms for photons, electrons, heavy-flavor jets, hadronic $\tau$ decays, missing transverse momentum, and particle flow will be essential for realizing the full physics potential of the detector. Ultimately, comprehensive studies of realistic signal and background processes, including beam-induced and collision-induced backgrounds, will be required to fully characterize the physics reach of a 10~TeV muon collider experiment.

\clearpage
\noindent
\textbf{Acknowledgments}

Thank you to the International Muon Collider Collaboration (IMCC) and the US Muon Collider Collaboration (USMCC) for fostering this effort.

Also thanks to Simone Pagan Griso, Matt LeBlanc, Jan Offermann, Alyna Tang, Jullian Watts, Devlin Jenkins, Ethan Martinez, Moses Glassman, Kevin Dewyspelaere, Giacomo Da Molin, Giovanni Battista Marozzo, and Michele Gallinaro for useful input, suggestions, and feedback.

This research was supported in part by grant NSF PHY-2309135 to the Kavli Institute for Theoretical Physics (KITP), where the idea was conceived and the team was put together.

This work was supported by the EU HORIZON Research and Innovation Actions under the grant agreement number 101094300.
Funded by the European Union (EU). Views and opinions expressed are however those of the author(s) only and do not necessarily reflect those of the EU or European Research Executive Agency (REA). Neither the EU nor the REA can be held responsible for them.

This work was supported by a grant from the Simons Foundation [SFI-MPS-T-MPS-00010555, I.O., T.R.H, K.F.D]. %

This work was supported by U.S. Department of Energy (DOE), Office of Science, Energy Frontier program under Award Number DE-SC0017647 (A.M., C.K.), Award Number DE-SC0020267 (T.R.H., L.L., M.H.), Award Number DE-SC0023122 (T.R.H.), and Award Number DE-SC0023321 (L.L.). L.L. and C.B. were additionally supported by the U.S. National Science Foundation (NSF) under Award No. 2235028 and financial support for T.R.H.'s group's contribution to this publication results from grant \#CS-CSA-2024-018 from Research Corporation for Science Advancement.

K.F.D., M.L., L.R, and B.R.'s work is supported by the Enrico Fermi Institute at the University of Chicago. K.F.D. was also supported by U.S NSF awards 2411692 and 2443370, as well as the Neubauer Family Assistant Professor Program. M.L. and B.R. were also supported by NSF award No. 2310094. 

R.P.'s work is supported by the National Science Foundation (Graduate Research Fellowship Program) under Grant No. DGE-2444107. Any opinions, findings, and conclusions or recommendations expressed in this material are those of the authors and do not necessarily reflect the views of the National Science Foundation.

F.M. acknowledges support by the Deutsche Forschungsgemeinschaft (DFG, German Research Foundation) under Germany’s Excellence Strategy – EXC 2121 ``Quantum Universe'' – 390833306.

This work has benefited from computing services provided by the German National Analysis Facility (NAF).
This research was conducted using services provided by the OSG Consortium~\cite{osg07,osg09,https://doi.org/10.21231/906p-4d78,https://doi.org/10.21231/0kvz-ve57}, which is supported by U.S. NSF awards 2030508 and 1836650. We are particularly grateful to Pascal Paschos for his support.

\bibliographystyle{bibstyle}
\bibliography{biblio}%

\end{document}